\documentclass[twocolumn,preprintnumbers,amsmath,amssymb]{revtex4}
\usepackage[utf8]{inputenc}
\usepackage{epsfig}
\usepackage{graphicx}
\usepackage{dcolumn}
\usepackage{bm}
\usepackage{amsmath, amsfonts}
\usepackage{amssymb}
\usepackage{amsthm}
\usepackage{physics}
\usepackage{comment}
\usepackage{color} 
\usepackage[colorlinks,linkcolor=blue,anchorcolor=blue,urlcolor=blue, citecolor=blue]{hyperref}
\usepackage{caption}
\usepackage{subcaption}
\usepackage{siunitx,booktabs}
\usepackage{epstopdf}
\usepackage{dsfont}

\begin{document}
	\title{Foiling zero-error attacks against coherent-one-way quantum key distribution}
	\author{Marcos Curty}
	\affiliation{Escuela de Ingeniería de Telecomunicación, Department of Signal Theory and Communications, ­University of Vigo, Vigo E-36310, Spain}
\begin{abstract}
To protect practical quantum key distribution (QKD) against photon-number-splitting attacks, one could measure the coherence of the received signals. One prominent example that follows this approach is coherent-one-way (COW) QKD, which is commercially available. Surprisingly, however, it has been shown very recently that its secret key rate scales quadratically with the channel transmittance, and, thus, this scheme is unsuitable for long-distance transmission. This result was derived by using a zero-error attack, which prevents the distribution of a secure key without introducing any error. Here, we study various countermeasures to foil zero-error attacks against COW-QKD. They require to either monitor some additional available detection statistics, or to increase the number of quantum states emitted. We obtain asymptotic upper security bounds on the secret key rate that scale close to linear with the channel transmittance, thus suggesting the effectiveness of the countermeasures to boost the performance of this protocol. 
\end{abstract}

\maketitle

\section{Introduction}

Quantum key distribution (QKD) allows two remote users (typically called Alice and Bob) to share a symmetric key with information-theoretic security~\cite{qkd1,qkd2,qkd3}. This key is the essential resource of the one-time-pad cryptosystem~\cite{pad}, the only known solution that guarantees confidential communication without resorting to computational complexity assumptions. 

Since its first theoretical proposal~\cite{prop1,prop2}, several QKD protocols have been introduced~\cite{prot1,prot1b,prot1c,cow1,prot2,prot3,prot4,prot5,prot6,prot7,prot8,prot9,prot10,prot11,prot12,prot13} and implemented~\cite{impl1,impl2,impl3,impl4,impl5,impl6,impl7,cow2,impl9,cow4,impl10,impl11,impl12,impl13} in the last two decades, each of them with their own merits. Recent experimental advances in QKD include, for instance, the deployment of QKD networks based on trusted nodes~\cite{net1,net2,net3,net4}, the distribution of secret keys over 600~km of optical fibre~\cite{impl14}, and the demonstration of satellite-based QKD links~\cite{sat1,sat2}.

Distributed-phase-reference (DPR) QKD~\cite{prot1b,prot1c,cow1,impl7,cow2,impl9,cow4} is a type of QKD protocols in which Alice and Bob check the coherence of the received optical pulses. This approach has triggered great attention in recent years because it held the promise to provide a secret key rate that scales linearly with the system's transmittance, $\eta_{\rm sys}$, by using a simple experimental setup. This is the best possible scaling for point-to-point QKD links~\cite{scale1,scale2}. DPR-QKD includes the commercially available~\cite{company} coherent-one-way (COW) QKD scheme~\cite{cow1,cow2,impl9}, which has been demonstrated over $300$~km of optical fibre~\cite{cow4}. Surprisingly, however, very recently it has been shown that COW-QKD provides a secret key rate that scales quadratically with $\eta_{\rm sys}$~\cite{upper,zero_cow}, which implies that all of its experimental implementations reported so far in the scientific literature are insecure. This matches the scaling of the lower security bounds against general attacks presented in~\cite{low_cow}, and renders COW-QKD unsuitable for long-distance QKD transmission. 

The results in~\cite{upper,zero_cow} use a special type of zero-error attack~\cite{cow_zer,cow_zer1}, in which the eavesdropper (Eve) measures out all the signals sent by Alice one by one, and then resends Bob new signals (whose state might depend on all her measurement results) that do not introduce any error. For this, Eve exploits two special properties of the signals emitted in COW-QKD. First, they are linearly independent, which means that they can be identified with an unambiguous state discrimination (USD) measurement~\cite{chefles_usd1,chefles_usd2,eldar1}. And, second, they include vacuum pulses, which naturally break the coherence between adjacent pulses. Such attack transforms the quantum channel into an entanglement breaking channel and, therefore, it does not allow the distribution of a secure key~\cite{condition}.  

In this paper, we evaluate the effectiveness of three possible countermeasures to foil zero-error attacks against COW-QKD and, thus, boost the performance of this protocol. For this, we consider the zero-error attack introduced in~\cite{zero_cow}, which is optimal when Eve measures Alice's signals one by one, and it outperforms previous approaches~\cite{upper,cow_zer,cow_zer1}. For each countermeasure analyzed, we derive asymptotic upper security bounds and compare our results with those corresponding to the original COW-QKD scheme. 

The first two countermeasures do not require to modify the experimental setup of COW-QKD. Instead, Alice and Bob need to monitor additional detection statistics of Alice's signals at Bob's side, besides the standard quantum bit error rate (QBER) and the visibilities prescribed in the original COW-QKD scheme. This includes the monitoring of the number of coincidences ({\it i.e.}, those detection events in which Bob observes a simultaneous ``click'' both in his data and monitoring lines), as well as the detection rates of Alice's signals at Bob's data line. This is so because zero-error attacks typically modify these quantities greatly, when compared to their expected values in the absence of Eve. The third countermeasure contemplates the scenario in which Alice increases the number of emitted signals by adding a vacuum signal~\cite{cow_zer}. In doing so, it now becomes harder for Eve to unambiguously identify the state of the signals emitted. We show that these countermeasures could be quite effective to protect COW-QKD against zero-error attacks.

The paper is organized as follows. In Sec.~\ref{cow}, we introduce COW-QKD. In Sec.~\ref{zero_attack}, we review briefly the optimal zero-error attack presented in~\cite{zero_cow}. Then, in Secs.~\ref{coincidences}, \ref{det_rates} and \ref{four}, we study each of the three countermeasures considered. Finally, Sec.~\ref{conclu} concludes the paper with a summary. The paper includes as well several Appendixes with additional calculations. 

\section{Coherent-one-way QKD}\label{cow}

\begin{figure}
\centering 
\centerline{\includegraphics*[scale=0.35]{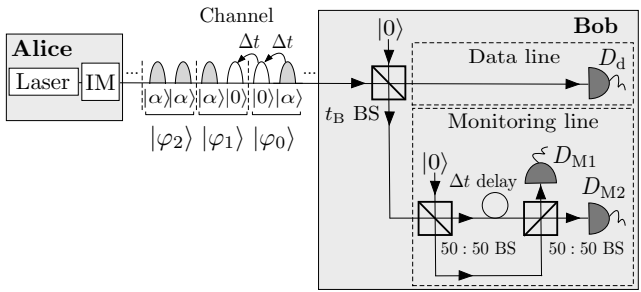}}
\caption{Alice uses a laser, together with an intensity modulator (IM), to randomly prepare the signals $\ket{\varphi_0}=\ket{0}\ket{\alpha}$, $\ket{\varphi_1}=\ket{\alpha}\ket{0}$ and $\ket{\varphi_2}=\ket{\alpha}\ket{\alpha}$. At Bob's side, a beamsplitter (BS) of transmittance $t_{\rm B}$ distributes the incoming signals between the data and the monitoring lines. The latter consists of a Mach-Zehnder interferometer. In the figure: $\ket{0}$ ($\ket{\alpha}$) is a vacuum state (coherent state of amplitude $\alpha$); $\Delta{}t$ denotes the time delay the two pulses within a signal; $D_{\rm d}$, $D_{\rm M_1}$ and $D_{\rm M_2}$ are single-photon detectors.}
\label{cow-setup}
\end{figure}
Fig.~\ref{cow-setup} shows the basic setup of the original COW-QKD scheme~\cite{cow1,cow2}. Alice sends Bob a sequence of signals $\ket{\varphi_0}=\ket{0}\ket{\alpha}$, $\ket{\varphi_1}=\ket{\alpha}\ket{0}$ and $\ket{\varphi_2}=\ket{\alpha}\ket{\alpha}$ that she selects at random each given time, where $\ket{0}$ ($\ket{\alpha}$) represents a vacuum (coherent) state. The signals $\ket{\varphi_i}$, with $i=0,1$, encode a bit value $i$ and are used for key generation. The decoy signal $\ket{\varphi_2}$, on the other hand, is used to test for eavesdropping. These signals are generated by Alice with {\it a priori} probabilities 
\begin{equation}\label{zxcv1}
p_{\ket{\varphi_0}}=p_{\ket{\varphi_1}}=\frac{1-f}{2}, \quad {\rm and}\quad p_{\ket{\varphi_2}}=f,
\end{equation}
with $0<f<1$.

At Bob's receiver, he uses a beamsplitter of transmittance $t_{\rm B}$ to randomly distribute the incoming signals into two possible lines: the data line and the monitoring line. The former measures the signals with a single-photon detector $D_{\rm d}$. Bob can distinguish the two key generation signals $\ket{\varphi_i}$, with $i=0,1$, based on the time slot in which $D_{\rm d}$ reports a detection ``click''. Precisely, if he observes a ``click'' in the first (second) time slot of a signal, then he considers that it is $\ket{\varphi_0}$ ($\ket{\varphi_1}$). If there is a double ``click'' ({\it i.e.}, a ``click'' in both time slots), Bob assigns a random bit value to it. The monitoring line tests the coherence between adjacent pulses. This is done by means of a Mach-Zehnder interferometer followed by two single-photon detectors $D_{\rm M1}$ and $D_{\rm M2}$. The interferometer is arranged such that say $D_{\rm M2}$ never produces a detection ``click'' when the state of the two adjacent pulses is $\ket{\alpha}$.

Once the quantum phase of the protocol ends, Bob announces over an authenticated classical channel which signals produced a ``click'' in his data line. Also, Alice announces which detected signals correspond to key generation signals (but without revealing their state $\ket{\varphi_i}$). The respective bit values associated to such signals form Alice and Bob's sifted key. In addition, they estimate the QBER in the sifted key, as well as the visibilities $V_{j}$ observed in the monitoring line. These visibilities are defined as
\begin{equation}
{V_j} = \frac{{p_{\rm click}({{\rm{D_{M1}}}|j}) - p_{\rm click}({{\rm{D_{M2}}}|j})}}{{p_{\rm click}({{\rm{D_{M1}}}|j}) + p_{\rm click}({{\rm{D_{M2}}}|j})}}, \label{vis_gen}
\end{equation}
where the index $j=\{``2$''$, ``01$''$, ``02$''$, ``21$''$, ``22$"$\}$ identifies the five possible signal combinations in which there are two adjacent coherent states $\ket{\alpha}$. For instance, $V_2$ refers to the interference between the two adjacent states $\ket{\alpha}$ in $\ket{\varphi_2}=\ket{\alpha}\ket{\alpha}$, $V_{01}$ considers the interference between the two adjacent states $\ket{\alpha}$ in $\ket{\varphi_0}\ket{\varphi_1}=\ket{0}\ket{\alpha}\ket{\alpha}\ket{0}$, and the other cases are defined similarly. In Eq.~(\ref{vis_gen}), $p_{\rm click}({\rm D}_{{\rm M}i}|j)$ denotes the conditional probability to observe a ``click'' in ${\rm D}_{{\rm M}i}$ that corresponds to the interference of two adjacent coherent states $\ket{\alpha}$ when they are situated within a signal combination $j$.

When the QBER (the visibilities $V_{j}$) is (are) above (below) a certain threshold, no secret key can be distilled from the sifted key~\cite{low_cow,cow4}, being the ideal noiseless case that where QBER$=0$ and  $V_{j}=1$ for all $j$. 

\section{Zero-error attack against COW-QKD}\label{zero_attack}

Here, we briefly review the zero-error attack introduced in~\cite{zero_cow}, which is optimal when Eve measures Alice's signals one by one. The key idea is rather simple. First, Eve measures out each of Alice's signals with an optimal USD measurement~\cite{chefles_usd1,chefles_usd2,eldar1}. Afterwards, she resends Bob all those blocks of signals that contain consecutive correctly identified signals, and whose first and last optical pulses are prepared in a vacuum state. In addition, Eve replaces the states $\ket{\alpha}$ within such blocks with coherent states $\ket{\beta}$ with $\beta\gg\alpha$. By selecting $\beta$ large enough, she can guarantee that each state $\ket{\beta}$ will produce a detection ``click'' at Bob's data line with basically unit probability. 

It is easy to show that this eavesdropping strategy does not introduce any error in Bob's data line nor it decreases the visibilities in his monitoring line. Indeed, since Eve's USD measurement never misidentifies Alice's signals, we have that QBER$=0$ and the visibilities related to the signals inside the blocks are perfect. Moreover, since the blocks start and end with correctly identified vacuum pulses, the visibilities remain perfect also in the borders of the blocks, independently of the signals that precede and follow them (and for which Eve obtained an inconclusive result with her USD measurement). In short, we have that $V_{j}=1$ for all $j$. 

Importantly, it can be shown that such zero-error attack is optimal in the sense that it maximises the gain $G_{\rm zero}$ at Bob's data line. This gain is defined as the probability that Bob observes a ``click'' in that line per signal sent by Alice. Remarkably, if the observed gain in an experimental implementation of COW-QKD exceeds $G_{\rm zero}$, Alice and Bob cannot generate a secure key. This is so because the observed data could be explained as coming from an entanglement breaking channel, and thus they do not share quantum correlations~\cite{condition}. 
 
\begin{figure}
\centering 
\centerline{\includegraphics*[scale=0.8]{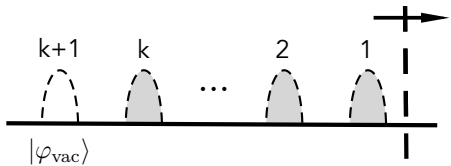}}
\caption{Block containing $k$ correctly identified signals $\ket{\varphi_i}$, with $i=0,1,2$, that Eve resends to Bob. These signals are illustrated in the figure with grey ovals (drawn with a dashed line). The white oval situated in the $k+1$th position of the block denotes a vacuum signal $\ket{\varphi_{\rm vac}}=\ket{0}\ket{0}$. This vacuum signal replaces the signal for which Eve obtained an inconclusive result with her USD measurement. The arrow indicates the direction of transmission towards Bob, and the dashed vertical line represents the beginning of the block.}
\label{fig:t2mmt}
\end{figure}

 The parameter $G_{\rm zero}$ can be written as~\cite{zero_cow}
\begin{eqnarray}\label{gzer}
G_{\rm zero}&=&\frac{1-p_{\rm c}}{1-p_{\rm c}^{M_{\rm max}+1}}\Bigg[\sum_{k=2}^{M_{\rm max}-1} p_{\rm c}^k(1-p_{\rm c}) p_{\rm click}(k) \nonumber \\
&+&p_{\rm c}^{M_{\rm max}}p_{\rm click}(M_{\rm max})\Bigg],
\end{eqnarray}
where $p_{\rm c}$ is the probability that Eve's USD measurement provides a conclusive result. The quantity $M_{\rm max}$ refers to the maximum length of the blocks of signals that Eve resends to Bob. In principle, $M_{\rm max}$ can be chosen arbitrary large. In practice, however, even relatively small values of $M_{\rm max}$ ({\it e.g.}, say $M_{\rm max}=10$) are sufficient to basically achieve the maximum possible value of $G_{\rm zero}$. This is because when the intensity of Alice's signals is small (as is the case in COW-QKD), the probability to obtain more than $M_{\rm max}$ consecutive conclusive measurement results is essentially negligible even for moderate values of $M_{\rm max}$. Finally, $p_{\rm click}(k)$ denotes the average number of ``clicks'' observed by Bob in his data line when Eve sends him a block containing $k$ correctly identified signals (see Fig.~\ref{fig:t2mmt}).

The quantity $p_{\rm click}(k)$ can be written as~\cite{zero_cow}
\begin{eqnarray}\label{pclickFirstSignalB}
p_{\rm click}(k)&=&\sum_{j=0}^2p(j|{\rm c})p_{\rm click}(k|j), 
\end{eqnarray}
where $p(j|{\rm c})$ denotes the conditional probability that Eve obtains the result $\ket{\varphi_j}$, with $j=0,1,2$, given that her USD measurement is conclusive, and $p_{\rm click}(k|j)$ refers to the average number of ``clicks'' observed by Bob in his data line when Eve sends him a block with $k$ correctly identified signals and the first signal of the block is in the state $\ket{\varphi_j}$.

The probabilities $p(j|{\rm c})$ are given by
\begin{eqnarray}\label{cond_prob_sun}
p(0|{\rm c})&=&p(1|{\rm c})=\frac{(1-f)q_{\rm s}}{2p_{\rm c}}, \nonumber \\
p(2|{\rm c})&=&1-p(0|{\rm c})-p(1|{\rm c})=1-\frac{(1-f)q_{\rm s}}{p_{\rm c}},
\end{eqnarray}
where $q_{\rm s}$ refers to the probability that Eve obtains a conclusive result when she measures a key generation signal $\ket{\varphi_j}$, with $j=0,1$. 

By calculating $p_{\rm click}(k|j)$ and explicitly computing Eq.~(\ref{pclickFirstSignalB}), it has been shown in~\cite{zero_cow} that $p_{\rm click}(k)$ can be expressed as
\begin{align}\label{FINAL_pClick_bb}
p_{\rm click}(k)&=\frac{1}{p(0|{\rm c})}\Big\{-1+(k+1) p(0|{\rm c})\\\nonumber&+\big[1-2p(0|{\rm c})\big]^k\big[1+(k-1)p(0|{\rm c})\big]\Big\},
\end{align}
for any $k\geq 2$. 

Most experimental implementations of COW-QKD satisfy
\begin{equation}\label{sun3}
\sqrt{\frac{f}{2(1-f)}}\leq e^{-\frac{1}{2}\mu},
\end{equation}
where $\mu=|\alpha|^2$ is the intensity of Alice's signals. In this scenario, the parameters $p_{\rm c}$ and $q_{\rm s}$ have the form~\cite{zero_cow} 
\begin{eqnarray}\label{sun2}
q_{\rm s}&=&1-e^{-\mu}, \nonumber \\
p_{\rm c}&=&(1-f)(1-e^{-\mu}).
\end{eqnarray}
By combining Eqs.~(\ref{cond_prob_sun})-(\ref{sun2}), we find that
\begin{equation}\label{dom}
p(0|{\rm c})=p(1|{\rm c})=\frac{1}{2}, \quad {\rm and}\quad p(2|{\rm c})=0.
\end{equation}
That is, Eve's optimal USD measurement does not identify decoy signals. Or to put it in other words, in this situation Eve resends Bob blocks of signals that contain only key generation signals $\ket{\varphi_i}$, with $i=0,1$. 

\section{Monitoring coincidences}\label{coincidences}

In this section, we study a first possible countermeasure against the zero-error attack described above. Precisely, this countermeasure exploits the fact that Eve sends Bob coherent states of very high intensity $\abs{\beta}^2$ (with $\beta\gg\alpha$) to maximise the gain $G_{\rm zero}$. In doing so, she can assure that the signals that are correctly identified by her USD measurement will produce a ``click'' at Bob's data line with basically unit probability. However, if $\abs{\beta}^2$ is too large, these signals will produce coincidence detection events. These are events in which Bob observes a simultaneous detection ``click'' (within the time duration of an optical pulse emitted by Alice) both in the data and monitoring lines. Therefore, Alice and Bob could monitor such coincidence detection events to detect Eve's zero-error attack. To put it in other words, to remain undetected, Eve should send weaker optical pulses to Bob, and thus $G_{\rm zero}$ will decrease. 

This countermeasure implicitly assumes the {\it trusted device scenario}, where Eve {\it cannot} modify the properties of Bob's measurement device. This might be a reasonable assumption for many practical situations, and is the scenario that we consider below. If Eve could change, say, the detection efficiency of Bob's detectors together with the transmittance $t_{\rm B}$, then monitoring coincidence detection events would not translate into relevant restrictions on her capabilities. This is so because, in that case, even single-photon pulses sent by Eve could produce a ``click'' at Bob's data line with essentially unit probability.

We shall define the coincidence gain of a zero-error attack, which we denote by $G_{\rm zero}^{\rm coin}$, as the probability that Bob observes a coincidence detection event per signal sent by Alice. By following the same procedure used in~\cite{zero_cow} to derive Eq.~(\ref{gzer}), it is straightforward to show that $G_{\rm zero}^{\rm coin}$ can be written as
\begin{eqnarray}\label{gcoin}
G_{\rm zero}^{\rm coin}&=&\frac{1-p_{\rm c}}{1-p_{\rm c}^{M_{\rm max}+1}}\Bigg[\sum_{k=2}^{M_{\rm max}-1} p_{\rm c}^k(1-p_{\rm c}) p_{\rm coin}(k) \nonumber \\
&+&p_{\rm c}^{M_{\rm max}}p_{\rm coin}(M_{\rm max})\Bigg],
\end{eqnarray}
where $p_{\rm coin}(k)$ represents the average number of coincidence detection events observed by Bob when Eve sends him a block containing $k$ correctly identified signals. This latter parameter can be expressed as
\begin{eqnarray}\label{pclickFirstSignalBB_or}
p_{\rm coin}(k)&=&\sum_{j=0}^2p(j|{\rm c})p_{\rm coin}(k|j), 
\end{eqnarray}
where $p_{\rm coin}(k|j)$ is the average number of coincidence detection events observed by Bob when Eve sends him a block containing $k$ correctly identified signals and the first signal of the block is in the state $\ket{\varphi_j}$. 

If the condition given by Eq.~(\ref{sun3}) holds, which is the experimental parameter regime in which we are interested, then from Eq.~(\ref{dom})-(\ref{pclickFirstSignalBB_or}) we have that
\begin{eqnarray}\label{pclickFirstSignalBB}
p_{\rm coin}(k)&=&\frac{1}{2}\left[p_{\rm coin}(k|0)+p_{\rm coin}(k|1)\right].
\end{eqnarray}

Similarly, from Eqs~(\ref{pclickFirstSignalB})-(\ref{dom}), we have that the parameter $p_{\rm click}(k)$ required to calculate $G_{\rm zero}$, can be written as
\begin{eqnarray}\label{pclickFirstSignalBBb}
p_{\rm click}(k)&=&\frac{1}{2}\left[p_{\rm click}(k|0)+p_{\rm click}(k|1)\right].
\end{eqnarray}
Here, we do not use Eq.~(\ref{FINAL_pClick_bb}) because it assumes that Eve sends Bob coherent states of very high intensity. However, as we show next, she could send him other signals to reduce the number of coincidence detection events. 

\subsection{Signals sent by Eve}

\begin{figure}
\centering 
\centerline{\includegraphics*[scale=0.8]{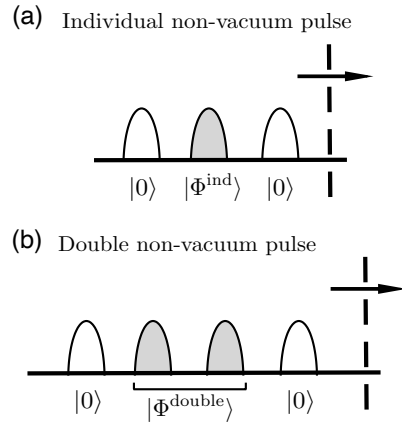}}
\caption{The non-vacuum optical pulses within a block of correctly identified signals that Eve resends to Bob are either surrounded by vacuum pulses (as shown in case (a) in the figure) or they have, at most, one adjacent non-vacuum pulse (as shown in case (b) in the figure). A grey (white) oval, drawn with a solid line, represents a non-vacuum (vacuum) optical pulse.}
\label{fig:t2}
\end{figure}

Due to the fact that Eve's optimal USD measurement does not identify decoy signals when Eq.~(\ref{sun3}) holds, all blocks of signals that Eve resends to Bob satisfy the following property: The non-vacuum optical pulses within the block are either surrounded by vacuum pulses or they have, at most, one adjacent non-vacuum pulse. This is illustrated in Fig.~\ref{fig:t2}. 

The former case arises, for instance, when Eve's correctly identified adjacent signals are in the same state $\ket{\varphi_j}$, with $j=0, 1$. We shall call these non-vacuum optical pulses as {\it individual}, and we will denote their quantum state as $\ket{\Phi^{\rm ind}}$. The latter case arises when Eve's correctly identified adjacent signals are in the state $\ket{\varphi_0}\ket{\varphi_1}$. We shall call these non-vacuum optical pulses as {\it double}, and we will denote their joint quantum state as $\ket{\Phi^{\rm double}}$.

We can always write the state $\ket{\Phi^{\rm ind}}$ as
\begin{equation}\label{f1n}
\ket{\Phi^{\rm ind}}=\sum_{n=0}^\infty q_n\ket{n},
\end{equation}
where $\sum_{n=0}^\infty \abs{q_n}^2=1$, and $\ket{n}$ is a Fock state with $n$ photons. That is, Eq.~(\ref{f1n}) simply expresses a general pure state $\ket{\Phi^{\rm ind}}$ in the Fock basis. 

In the case of {\it double} non-vacuum optical pulses, Eve must preserve the coherence between the two pulses. The mode that preserves such coherence has the form~\cite{cow_zer}
\begin{equation}
a_{12}^\dagger=\frac{a_1^\dagger+a_2^\dagger}{\sqrt{2}}
\end{equation}
where $a_i^\dagger$, with $i=1,2$, denotes the creation operator of photons in the time instants associated to the two non-vacuum pulses. This means, in particular, that $\ket{\Phi^{\rm double}}$ can be expressed as
\begin{equation}\label{Mon3}
\ket{\Phi^{\rm double}}=\sum_{n=0}^\infty p_n\ket{n},
\end{equation}
where $\sum_{n=0}^\infty \abs{p_n}^2=1$ and $\ket{n}=(a_{12}^\dagger)^n/\sqrt{n!}\ket{0}$ is a Fock state with $n$ photons in total in the two non-vacuum pulses. 

In the next section we calculate the parameters $p_{\rm click}(k|j)$ and $p_{\rm coin}(k|j)$, with $j=0,1$, that are needed to evaluate Eqs.~(\ref{pclickFirstSignalBB})-(\ref{pclickFirstSignalBBb}).

\subsection{Probabilities $p_{\rm click}(k|j)$ and $p_{\rm coin}(k|j)$}

For simplicity, in the calculations below, and also in other sections of this paper, we shall disregard the effect of the dark counts of Bob's detectors. 

\subsubsection{Probabilities $p_{\rm click}(k|1)$ and $p_{\rm coin}(k|1)$}\label{qwe1}

\begin{figure}
\centering 
\centerline{\includegraphics*[scale=0.7]{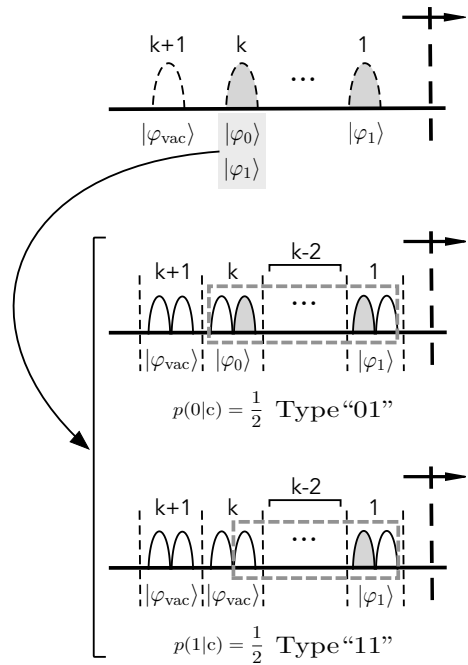}}
\caption{Schematic representation of a block containing $k$ correctly identified signals by Eve, and where the first signal of the block is $\ket{\varphi_1}$. Here, each big oval (drawn with a dashed line) represents a signal ({\it i.e.}, it contains two optical pulses), while small ovals (drawn with a solid line) denote optical pulses. The signals in the first $k$ positions of the block correspond to the correctly identified signals, while the signal located in the $k+1$th position of the block is a vacuum signal $\ket{\varphi_{\rm vac}}=\ket{0}\ket{0}$. There are two options, depending on the state of the signal located in the $k$th position. With probability $p(0|{\rm c})=1/2$ ($p(1|{\rm c})=1/2$) this signal is $\ket{\varphi_0}$ ($\ket{\varphi_1}$). If it is $\ket{\varphi_1}$, Eve replaces it with a vacuum signal $\ket{\varphi_{\rm vac}}$ to guarantee that the resulting sub-block starts and ends with correctly identified vacuum {\it pulses}. The sub-blocks of signals that Eve resends to Bob are illustrated with grey dashed rectangles.
}
\label{fig:t3a}
\end{figure}

This scenario is illustrated in Fig.~\ref{fig:t3a}. 
Since the first signal of the block is $\ket{\varphi_1}$, and the first optical pulse of this signal is a vacuum pulse, the longest sub-block that starts and ends with vacuum pulses (if there is any) must include this signal $\ket{\varphi_1}$. Now, there are two options, depending on the state of the signal located in the $k$th position of the block. If this signal is $\ket{\varphi_0}$, which happens with probability $p(0|{\rm c})=1/2$, then the block starts and ends with vacuum pulses, and Eve sends this block to Bob. We shall call this block as a sub-block of the type ``01''. It is illustrated in Fig.~\ref{fig:t3a} with a grey dashed rectangle. On the other hand, if the signal in the $k$th position of the block is $\ket{\varphi_1}$, which happens with probability $p(1|{\rm c})=1/2$, then the block ends with a non-vacuum optical pulse. Therefore, Eve replaces the signal $\ket{\varphi_1}$ with a vacuum signal $\ket{\varphi_{\rm vac}}$. In doing so, she guarantees that the block now ends with a correctly identified vacuum pulse (which corresponds to the first optical vacuum pulse in $\ket{\varphi_1}$). We shall refer to this sub-block as being of the type ``11'', and is illustrated in Fig.~\ref{fig:t3a} also with a grey dashed rectangle. 

We define the parameters $N_{01}^{\rm ind}(k)$ ($N_{01}^{\rm double}(k)$) as the average number of {\it individual} ({\it double}) non-vacuum optical pulses (see Fig.~\ref{fig:t2}) contained in a sub-block of signals of the type ``01'' that Eve obtains from a block with $k$ correctly identified signals. Likewise, we define the analogous parameters $N_{11}^{\rm ind}(k)$ and $N_{11}^{\rm double}(k)$ for the sub-blocks of signals of the type ``11''. These parameters are calculated in Appendix~\ref{apA}, and their values are provided in Table~\ref{table1nnn}. 
\begin{table}[h!]
  \centering
  \begin{tabular}{ll}
    \hline\hline
 $N_{00}^{\rm ind}(2)=1$  \quad\quad\quad\quad & \quad $N_{00}^{\rm ind}(k)=\frac{k-1}{2}$ $\forall k\geq{}3$\\
 \hline
$N_{01}^{\rm ind}(2)=0$  \quad\quad\quad\quad & \quad $N_{01}^{\rm ind}(k)=\frac{k-1}{2}$ $\forall k\geq{}3$\\
  \hline
$N_{10}^{\rm ind}(2)=0$  \quad\quad\quad\quad & \quad $N_{10}^{\rm ind}(k)=\frac{k-1}{2}$ $\forall k\geq{}3$\\
 \hline
$N_{11}^{\rm ind}(2)=1$  \quad\quad\quad\quad & \quad $N_{11}^{\rm ind}(k)=\frac{k-1}{2}$ $\forall k\geq{}3$\\
\hline
$N_{00}^{\rm double}(2)=0$  \quad & \quad $N_{00}^{\rm double}(k)=\frac{k-1}{4}$ $\forall k\geq{}3$\\
\hline
$N_{01}^{\rm double}(2)=1$  \quad & \quad $N_{01}^{\rm double}(k)=\frac{k+1}{4}$ $\forall k\geq{}3$\\
\hline
$N_{10}^{\rm double}(2)=0$  \quad & \quad $N_{10}^{\rm double}(k)=\frac{k-3}{4}$ $\forall k\geq{}3$\\
\hline
$N_{11}^{\rm double}(2)=0$  \quad & \quad $N_{11}^{\rm double}(k)=\frac{k-1}{4}$ $\forall k\geq{}3$\\
 \hline\hline
\end{tabular}
\caption{Values of the parameters $N_{ij}^{\rm ind}(k)$ and $N_{ij}^{\rm double}(k)$, with $i,j=0,1$. $N_{ij}^{\rm ind}(k)$ ($N_{ij}^{\rm double}(k)$) represents the average number of {\it individual} ({\it double}) non-vacuum optical pulses contained in a sub-block of signals of the type ``ij'' that Eve obtains from a block with $k$ correctly identified signals.
  }\label{table1nnn}
\end{table}

Putting all together, we find that the probability $p_{\rm click}(k|1)$ can be written as
\begin{eqnarray}\label{as4}
p_{\rm click}(k|1)&=&p(0|{\rm c})\left[N_{01}^{\rm ind}(k)P_{\rm click}^{\rm ind}+N_{01}^{\rm double}(k)P_{\rm click}^{\rm double}\right]\nonumber \\
&+&
p(1|{\rm c})\left[N_{11}^{\rm ind}(k)P_{\rm click}^{\rm ind}+N_{11}^{\rm double}(k)P_{\rm click}^{\rm double}\right]\nonumber \\
&=&\frac{1}{2}\Big\{\left[N_{01}^{\rm ind}(k)+N_{11}^{\rm ind}(k)\right]P_{\rm click}^{\rm ind}\nonumber \\
&+&\left[N_{01}^{\rm double}(k)+N_{11}^{\rm double}(k)\right]P_{\rm click}^{\rm double}\Big\},
\end{eqnarray}
where in the second equality we have used the fact that $p(0|{\rm c})=p(1|{\rm c})=\frac{1}{2}$. The quantity $P_{\rm click}^{\rm ind}$ ($P_{\rm click}^{\rm double}$) denotes the average number of ``clicks'' observed by Bob in his data line when Eve sends him an {\it individual} ({\it double}) non-vacuum optical pulse. These quantities can be calculated from the form of the quantum states, $\ket{\Phi^{\rm ind}}$ and $\ket{\Phi^{\rm double}}$, given, respectively, by Eqs.~(\ref{f1n})-(\ref{Mon3}). 

Similarly, it is straightforward to show that the parameter $p_{\rm coin}(k|1)$ can be written as
\begin{eqnarray}\label{parque1}
p_{\rm coin}(k|1)&=&\frac{1}{2}\Big\{\left[N_{01}^{\rm ind}(k)+N_{11}^{\rm ind}(k)\right]P_{\rm coin}^{\rm ind}\nonumber \\
&+&\left[N_{01}^{\rm double}(k)+N_{11}^{\rm double}(k)\right]P_{\rm coin}^{\rm double}\Big\},\  \ \ \ \ 
\end{eqnarray}
where $P_{\rm coin}^{\rm ind}$ ($P_{\rm coin}^{\rm double}$) represents the average number of coincidence detection events observed by Bob when Eve sends him an {\it individual} ({\it double}) non-vacuum optical pulse. 

The parameters $P_{\rm click}^{\rm ind}$, $P_{\rm click}^{\rm double}$, $P_{\rm coin}^{\rm ind}$ and $P_{\rm coin}^{\rm double}$ are calculated in Sec.~\ref{qwe}.  

\subsubsection{Probabilities $p_{\rm click}(k|0)$ and $p_{\rm coin}(k|0)$} 

\begin{figure}
\centering 
\centerline{\includegraphics*[scale=0.7]{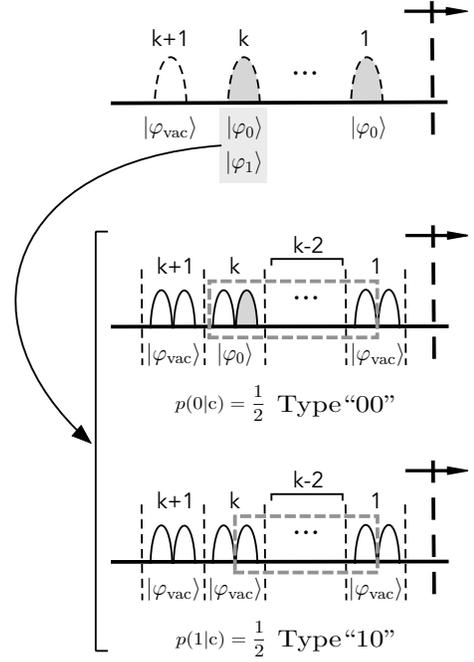}}
\caption{Schematic representation of a block containing $k$ correctly identified signals by Eve, and where the first signal of the block is $\ket{\varphi_0}$. Like in Fig.~\ref{fig:t3a}, each big oval (drawn with a dashed line) represents a signal ({\it i.e.}, it contains two optical pulses), while small ovals (drawn with a solid line) denote optical pulses. The signals in the first $k$ positions of the block correspond to the correctly identified signals, while the signal located in the $k+1$th position of the block is a vacuum signal $\ket{\varphi_{\rm vac}}=\ket{0}\ket{0}$. Eve replaces the signal $\ket{\varphi_0}$ in the first position of the block with a signal $\ket{\varphi_{\rm vac}}$ to guarantee that the resulting sub-block starts with a correctly identified vacuum {\it pulse}. Now, there are two options, depending on the state of the signal located in the $k$th position. With probability $p(0|{\rm c})=1/2$ ($p(1|{\rm c})=1/2$) this signal is $\ket{\varphi_0}$ ($\ket{\varphi_1}$). If it is $\ket{\varphi_1}$, Eve also replaces this signal with a vacuum signal $\ket{\varphi_{\rm vac}}$ to guarantee that the resulting sub-block ends with a correctly identified vacuum {\it pulse}. The sub-blocks of signals that Eve resends to Bob are illustrated with grey dashed rectangles.
}
\label{fig:t3b}
\end{figure}

The procedure to obtain $p_{\rm click}(k|0)$ and $p_{\rm coin}(k|0)$ is completely analogous to that used to calculate $p_{\rm click}(k|1)$ and $p_{\rm coin}(k|1)$ in Sec.~\ref{qwe1}, and, for simplicity, we omit the details here. This scenario is illustrated in Fig.~\ref{fig:t3b}. We find that
\begin{eqnarray}\label{as5}
p_{\rm click}(k|0)&=&\frac{1}{2}\Big\{\left[N_{10}^{\rm ind}(k)+N_{00}^{\rm ind}(k)\right]P_{\rm click}^{\rm ind}\nonumber \\
&+&\left[N_{10}^{\rm double}(k)+N_{00}^{\rm double}(k)\right]P_{\rm click}^{\rm double}\Big\}, \nonumber \\
p_{\rm coin}(k|0)&=&\frac{1}{2}\Big\{\left[N_{10}^{\rm ind}(k)+N_{00}^{\rm ind}(k)\right]P_{\rm coin}^{\rm ind}\nonumber \\
&+&\left[N_{10}^{\rm double}(k)+N_{00}^{\rm double}(k)\right]P_{\rm coin}^{\rm double}\Big\},\ \ \
\end{eqnarray}
where the parameters $N_{00}^{\rm ind}(k)$ and $N_{00}^{\rm double}(k)$ ($N_{10}^{\rm ind}(k)$ and $N_{10}^{\rm double}(k)$) refer to the sub-blocks of the type ``00'' (``10'') illustrated in Fig.~\ref{fig:t3b}. Their values are also provided in Table~\ref{table1nnn}.

\subsection{Parameters $P_{\rm click}^{\rm ind}$, $P_{\rm click}^{\rm double}$, $P_{\rm coin}^{\rm ind}$ and $P_{\rm coin}^{\rm double}$}\label{qwe}

It can be shown (see Appendix~\ref{qwe_ap}) that when the state $\ket{\Phi^{\rm ind}}$ given by Eq.~(\ref{f1n}) enters Bob's receiver surrounded by vacuum pulses, the parameters $P_{\rm click}^{\rm ind}$ and $P_{\rm coin}^{\rm ind}$ have the form
\begin{eqnarray}\label{resA}
P_{\rm click}^{\rm ind}&=&1-\sum_{{\it n}=0}^\infty \abs{{\it q}_{\it n}}^2\left(1-t_{\rm B}\eta_{\rm det}\right)^{\it n}, \nonumber \\
P_{\rm coin}^{\rm ind}&=&1-\sum_{n=0}^\infty \abs{q_n}^2\Big\{\left[1-(1-t_{\rm B})\frac{\eta_{\rm det}}{2}\right]^n\nonumber \\
&+&\left(1-\eta_{\rm det}t_{\rm B}\right)^n-\left[1-(1+t_{\rm B})\frac{\eta_{\rm det}}{2}\right]^n \Big\}.\ \ \ 
\end{eqnarray}

Similarly, when the state $\ket{\Phi^{\rm double}}$ given by Eq.~(\ref{Mon3}) enters Bob's receiver surrounded by vacuum pulses, the parameters $P_{\rm click}^{\rm double}$ and $P_{\rm coin}^{\rm double}$ have the form
\begin{eqnarray}\label{resB}
P_{\rm click}^{\rm double}&=&2\left[1-\sum_{n=0}^\infty \abs{p_n}^2\left(1-\frac{t_{\rm B}\eta_{\rm det}}{2}\right)^n\right], \nonumber \\
P_{\rm coin}^{\rm double}&=&2+\sum_{n=0}^\infty \abs{p_n}^2\bigg\{\left[1-\frac{\eta_{\rm det}}{2}\right]^n-2\left[1-\frac{t_{\rm B}\eta_{\rm det}}{2}\right]^n\nonumber \\
&+&\left[1-\frac{(1+t_{\rm B})\eta_{\rm det}}{4}\right]^n-\left[1-\frac{(1-t_{\rm B})\eta_{\rm det}}{2}\right]^n\nonumber \\
&-&\left[1-\frac{(1-t_{\rm B})\eta_{\rm det}}{4}\right]^n\bigg\}.
\end{eqnarray}
The calculations are provided in Appendix~\ref{qwe_ap2}.

\subsection{Parameters $p_{\rm click}(k)$ and $p_{\rm coin}(k)$}

By combining Eqs.~(\ref{pclickFirstSignalBB})-(\ref{pclickFirstSignalBBb})-(\ref{as4})-(\ref{parque1})-(\ref{as5}), and substituting the values of the parameters $N_{ij}^{\rm ind}(k)$ and $N_{ij}^{\rm double}(k)$, with $i,j=0,1$, given by Table~\ref{table1nnn}, we obtain that $p_{\tau}(k)$, with $\tau\in\{{\rm click}, {\rm coin}\}$, can be expressed as
\begin{eqnarray}\label{as7}
p_{\tau}(k)=\frac{k-1}{2}\left\{P_{\tau}^{\rm ind}+\frac{P_{\tau}^{\rm double}}{2}\right\}, 
\end{eqnarray}
for $k\geq{}2$. If we further insert in this equation (see Appendix~\ref{parque2}) the values of $P_{\tau}^{\rm ind}$ and $P_{\tau}^{\rm double}$ given by Eqs.~(\ref{resA})-(\ref{resB}), we find that $p_{\tau}(k)$ has the form given by Eq.(\ref{as6}).

\subsection{Evaluation}

To evaluate the effectiveness of this countermeasure, we optimize numerically the probability distributions $\abs{q_n}^2$ and $\abs{p_n}^2$ corresponding to the signal states sent by Eve, such that, for each achievable value of the gain $G_{\rm zero}$, they provide the minimum value of the coincidence detection rate $G_{\rm zero}^{\rm coin}$. 

For simplicity, in the simulations below, we consider that Eve's signals contain at most five photons ({\it i.e.}, $q_n=p_n=0$ for $n>5$), which is the regime that is relevant for the experimental values considered, which are provided in Table~\ref{tab:korzhParams}. They correspond to those COW-QKD implementations reported in~\cite{cow4} in which Alice uses the highest and the lowest intensity value $\mu=\abs{\alpha}^2$ for her signals. Moreover, we fix the parameter $M_{\rm max}=10$ and, as already mentioned, we disregard the effect of the dark counts of Bob's detectors.   
\begin{table}
\centering
\begin{tabular}{ |c|c|c|c|c|c|c| } 
 \hline
 $\mu$ & Att. [dB] & Distance (km) &  $\alpha_{\rm channel}$ & $\eta_{\rm det}$  & $f$ & $t_{\rm B}$\\ 
 \hline
 0.06  & 16.9 & 104  & 0.1625 & 0.22 & 0.155 & 0.9 \\ 
 \hline
 0.1   & 34.1 &  203 & 0.1680 & 0.27 & 0.155 & 0.9 \\ 
 \hline
\end{tabular}
\caption{Experimental parameters corresponding to those COW-QKD demonstrations reported in~\cite{cow4} that use the highest and the lowest intensity value $\mu=|\alpha|^2$ for Alice's signals. The parameter ``Att.'' in the table refers to the attenuation due only to channel loss, the distance refers to the fibre length, $\alpha_{\rm channel}$ represents the loss coefficient of the channel, $\eta_{\rm det}$ is the efficiency of Bob's detectors, $f$ denotes the probability that Alice emits a decoy signal $\ket{\varphi_2}$, and $t_{\rm B}$ is the transmittance of Bob's beamsplitter.}\label{tab:korzhParams}
\end{table}

\begin{figure}
\centering 
\centerline{\includegraphics*[scale=0.42]{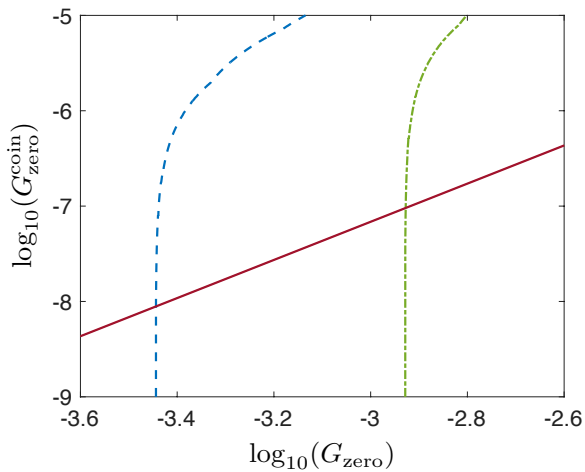}}
\caption{Minimum achievable values of $G_{\rm zero}^{\rm coin}$ as a function of $G_{\rm zero}$ in logarithmic scale for the experimental parameters provided in Table~\ref{tab:korzhParams}. The dashed blue line (dash-dotted green line) corresponds to the case $\mu=0.06$ ($\mu=0.1$) shown in that table. The solid magenta line illustrates the expected value of the coincidence detection rate as a function of the gain in the absence of Eve for a typical channel model. In order for Eve to remain undetected, $G_{\rm zero}^{\rm coin}$ must match the expected value given by the solid magenta line. This reduces the maximum possible value of $G_{\rm zero}$ when compared to the results provided in \cite{zero_cow}. See the text for further details.}
\label{fig:t2bbc}
\end{figure}

The results are illustrated in Fig.~\ref{fig:t2bbc}. The dashed blue line (dash-dotted green line) shows the minimum value of $G_{\rm zero}^{\rm coin}$ in logarithmic scale achievable by Eve as a function of $G_{\rm zero}$ for the experimental parameters corresponding to $\mu=0.06$ ($\mu=0.1$) in Table~\ref{tab:korzhParams}. When one increases the probability that Eve sends Bob higher photon number states, the values of $G_{\rm zero}$ and $G_{\rm zero}^{\rm coin}$ increase, and eventually $G_{\rm zero}$ matches the results obtained in~\cite{zero_cow}, though this is not shown in the figure. The solid magenta line corresponds to the expected coincidence detection rate as a function of the gain in the absence of Eve for a typical channel model. These expected values are calculated in Appendix~\ref{exp}. Precisely, the expected coincidence detection rate, which we denote by $G_{\rm coin}$, is given by Eq.~(\ref{f4}), while the expected gain, which we denote by $G$, is given by Eq.~(\ref{f3}). In reality, in Fig.~\ref{fig:t2bbc} there are two magenta lines, corresponding to the two sets of experimental parameters provided in Table~\ref{tab:korzhParams}. However, both lines essentially overlap each other and cannot be distinguished with the resolution of the figure. 

To remain undetected, Eve must guarantee that $G_{\rm zero}^{\rm coin}=G_{\rm coin}$. This reduces the maximum value of $G_{\rm zero}$ for which Eve's attack is successful, as shown in Fig.~\ref{fig:t2bbc}. The values of $G_{\rm zero}$ that satisfy this latter condition are provided in Table~\ref{tab:comparison}. According to these results, by monitoring coincidences Alice and Bob can approximately double the maximum achievable distance, which we denote by $L_{\rm zero}$, when compared to~\cite{zero_cow}. That is, $L_{\rm zero}$ corresponds to the transmission distance that provides a gain at Bob's side equal to $G_{\rm zero}$ in the absence of Eve's attack, for the channel model considered in Appendix~\ref{exp} and the experimental parameters provided in~Table~\ref{tab:korzhParams}. 
\begin{table}
\centering
\begin{tabular}{ |l|c|l|c| } 
 \hline
       & $\log_{10}(G_{\rm zero})$ & Att. [dB] & $L_{\rm zero}$(km)  \\ 
 \hline
 This work $\mu=0.06$                  & -3.44  & $\ \ \approx 15.8$ & $\approx 97$   \\        
 \hline
 Attack in~\cite{zero_cow} $\mu=0.06$ & -2.62 & $\ \ \approx 7.6$ &   $\approx 47$  \\ 
  \hline
This work $\mu=0.1$                  & -2.92  & $\ \ \approx 13.7$ & $\approx 81$  \\ 
 \hline
Attack in~\cite{zero_cow} $\mu=0.1$ & -2.19 & $\ \ \approx 6.4$ &  $\approx 38$  \\ 
 \hline
\end{tabular}
\caption{Values of $G_{\rm zero}$ that satisfy $G_{\rm zero}^{\rm coin}=G_{\rm coin}$. The parameter ``Att.'' ($L_{\rm zero}$) refers to the channel loss (distance) associated to $G_{\rm zero}$ if one considers the channel model described in Appendix~\ref{exp} and the experimental parameters provided in Table~\ref{tab:korzhParams}. For comparison, this table also includes the maximum values of $G_{\rm zero}$ obtained in~\cite{zero_cow} when one does not impose any restriction on $G_{\rm zero}^{\rm coin}$.}\label{tab:comparison}
\end{table}

While this is a positive result, the robustness of COW-QKD against channel loss still remains quite limited even if coincidence detection events are monitored, specially when compared to protocols like decoy-state QKD~\cite{prot2, prot3, prot4, impl1, impl2, impl3,impl4}. Indeed, the resulting values of $G_{\rm zero}$ are very close to those that could be obtained if Eve mainly sends single-photon pulses to Bob ({\it i.e.}, when $\abs{q_1}^2=\abs{p_1}^2=1$). As we will see in the next two sections, the other two countermeasures considered turn out to be significantly more effective to protect COW-QKD against zero-error attacks. Moreover, in their analysis we assume the conservative {\it untrusted device scenario}.

\section{Monitoring detection rates}\label{det_rates}

Here, we consider another possible countermeasure against the zero-error attack presented in Sec.~\ref{zero_attack}. It exploits the fact that when Eq.~(\ref{sun3}) holds, which happens in most implementations of COW-QKD that satisfy $f<0.5$ and $\mu<0.5$, Eve's optimal USD measurement does not identify decoy signals. Therefore, these signals are never forwarded to Bob. This means that Alice and Bob could monitor the detection rates of the emitted signals to detect a zero-error attack. Indeed, this approach has been considered in~\cite{cow_zer} against a restricted class of this type of attack. 

To remain undetected, Eve will have to modify her attack to reproduce the expected detection rates of the signals. Importantly, this implies the following. First, she will have to use a sub-optimal USD measurement capable of identifying decoy signals. And, second, she will have to post-process the blocks of correctly identified signals by using a sub-optimal strategy. This is so because the post-processing strategy in~\cite{zero_cow} favours the transmission of key generation signals. Indeed, to keep all the visibilities observed by Bob equal to one, the blocks of signals that Eve resends him cannot include decoy signals in their edges. For instance, in the extreme case of blocks of length $k=2$ ({\it i.e.}, blocks that contain precisely two correctly identified signals), which, on the other hand, have the highest probability to occur, none of the signals within the block can be a decoy signal. Thus, to match the expected detection rate of these signals, Eve will have to discard some blocks depending on their length.

Precisely, we define the gain of the decoy signals in a zero-error attack, which we shall denote by $G_{\rm zero}^{\rm decoy}$, as the probability that Bob observes a detection ``click'' in his data due to these signals. By using exactly the same procedure employed to obtain Eqs.~(\ref{gzer})-(\ref{gcoin}), it is straightforward to show that $G_{\rm zero}^{\rm decoy}$ is given by
\begin{eqnarray}\label{gdecoy}
G_{\rm zero}^{\rm decoy}&=&\frac{1-p_{{\rm c},\zeta}}{1-p_{{\rm c},\zeta}^{M_{\rm max}+1}}\Bigg[\sum_{k=2}^{M_{\rm max}-1} \gamma_k p_{{\rm c},\zeta}^k(1-p_{{\rm c},\zeta}) p_{\rm click}^{\rm decoy}(k) \nonumber \\
&+&\gamma_{M_{\rm max}}p_{{\rm c},\zeta}^{M_{\rm max}}p_{\rm click}^{\rm decoy}(M_{\rm max})\Bigg].
\end{eqnarray}
Here, $p_{{\rm c},\zeta}$ denotes the probability that Eve's USD measurement provides a conclusive outcome. This quantity now depends on a parameter $\zeta\in[f,1]$, which allows Eve to adjust the probability to correctly identify a decoy signal, between zero (when $\zeta=f$~\cite{foot2}) and its maximum allowed value (when $\zeta=1$). For further details see Appendix~\ref{meas_fri}. The probability $p_{\rm click}^{\rm decoy}(k)$, on the other hand, represents the average number of ``clicks'' observed by Bob in his data line due to a decoy signal when Eve sends him a block containing $k$ correctly identified signals. This quantity is given by
\begin{eqnarray}\label{pclickFirstSignalBB_B}
p_{\rm click}^{\rm decoy}(k)&=&\sum_{j=0}^2p(j|{\rm c})p_{\rm click}^{\rm decoy}(k|j),
\end{eqnarray}
where $p_{\rm click}^{\rm decoy}(k|j)$ is the average number of ``clicks'' observed by Bob in his data line due to a decoy signal when Eve sends him a block containing $k$ correctly identified signals and the first signal of the block is $\ket{\varphi_j}$. 

We note that in Eq.~(\ref{gdecoy}), we have also included certain parameters $\gamma_k\in[0,1]$, with $k=2,\ldots, M_{\rm max}$. They denote the probability that Eve actually resends Bob a block that contains $k$ correctly identified signals. This means that, in this scenario, the gain $G_{\rm zero}$ now reads
\begin{eqnarray}\label{gzer_decoy}
G_{\rm zero}&=&\frac{1-p_{{\rm c},\zeta}}{1-p_{{\rm c},\zeta}^{M_{\rm max}+1}}\Bigg[\sum_{k=2}^{M_{\rm max}-1} \gamma_k p_{{\rm c},\zeta}^k(1-p_{{\rm c},\zeta}) p_{\rm click}(k) \nonumber \\
&+&\gamma_{M_{\rm max}}p_{{\rm c},\zeta}^{M_{\rm max}}p_{\rm click}(M_{\rm max})\Bigg],
\end{eqnarray}
where the parameter $p_{\rm click}(k)$ is given by Eq.~(\ref{FINAL_pClick_bb}).

We shall denote the expected values of the analogous parameters in the absence of Eve for a typical channel model as $G_{\rm decoy}$ and $G$. That is, $G_{\rm decoy}$ ($G$) is the expected detection rate of the decoy signals (expected gain). These two quantities are calculated in Appendixes~\ref{exp2} and~\ref{exp}, respectively. This means that for $G_{\rm zero}=G$, Eve must guarantee that $G_{\rm zero}^{\rm decoy}=G_{\rm decoy}$. Importantly, since the expected gain of the key generation signals is equal for both of them, and this property is preserved by the zero-error attack, this condition is sufficient to assure that the detection rates of all the signals sent by Alice match their expected values. 

In the next section, we compute $p_{\rm click}^{\rm decoy}(k|j)$, with $j=0,1,2$. Eve's USD measurement, including the value of the parameter $p_{{\rm c},\zeta}$, is described in Appendix~\ref{meas_fri}.

\subsection{Parameters $p_{\rm click}^{\rm decoy}(k|j)$}

In the calculations below, we assume that the signals resent by Eve have sufficient intensity to produce a detection ``click'' at Bob's side with basically unit probability. For instance, Eve can resend Bob coherent states of very high intensity. 

\subsubsection{Parameter $p_{\rm click}^{\rm decoy}(k|1)$}

\begin{figure}
\centering 
\centerline{\includegraphics*[scale=0.6]{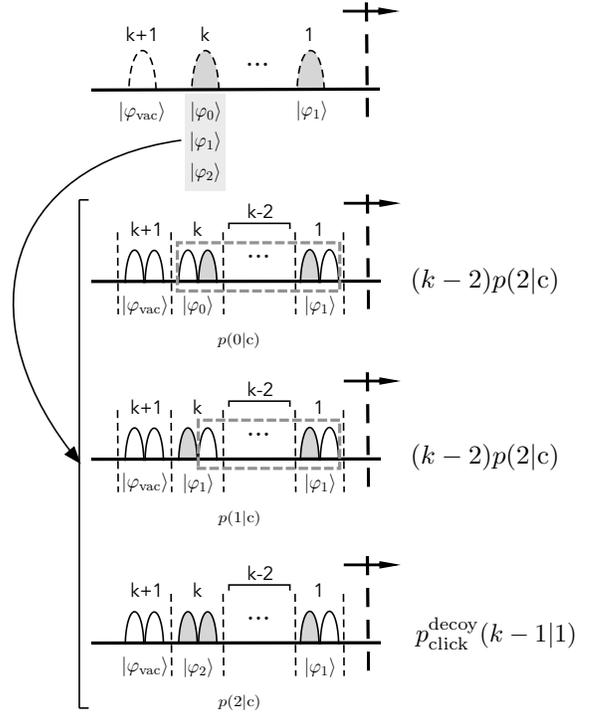}}
\caption{Illustration of the three sub-cases that we consider to evaluate $p_{\rm click}^{\rm decoy}(k|1)$. With probability $p(0|{\rm c})$ the signal in the $k$-th position of the block is $\ket{\varphi_0}$. In this case, when Eve resends Bob all the $k$ conclusive signals within the block (as it starts and ends with correctly identified vacuum pulses). The number of ``clicks'' at Bob's data line due to a decoy signal is $(k-2)p(2|{\rm c})$. This is so because decoy signals can only be located in the $k-2$ positions within the block ({\it i.e.}, they cannot be located in the edges of the block). The other two sub-cases are analogous and are described in the text. In the figure, large ovals (drawn with a dashed line) represent signals, while small ovals (drawn with a solid line) represent optical pulses within a signal. The sub-blocks of signals that Eve resends to Bob are illustrated with grey dashed rectangles.
}
\label{fig:decoy1}
\end{figure}

This scenario is illustrated in Fig.~\ref{fig:decoy1}.
Since the first optical pulse of $\ket{\varphi_1}$ is a vacuum pulse, the longest sub-block that starts and ends with correctly identified vacuum {\it pulses} (if there is any) includes $\ket{\varphi_1}$. 

With probability $p(0|{\rm c})$ the signal in the $k$-th position of the block is $\ket{\varphi_0}$. This means that Eve resends Bob all the $k$ signals in the block, as it starts and ends with correctly identified vacuum pulses. The decoy signals can only be located from position $2$ to position $k-1$ within the block. Therefore, Bob will observe on average $(k-2)p(2|{\rm c})$ ``clicks'' from decoy signals. 

Similarly, if the $k$-th signal of the block is $\ket{\varphi_1}$, which happens with probability $p(1|{\rm c})$, Eve replaces that signal with $\ket{\varphi_{\rm vac}}$ (not shown in the figure) and she resends Bob the first $k-1$ conclusive signals in the block because such sub-block has correctly identified vacuum pulses on its edges. Again, decoy signals can only be located from position $2$ to position $k-1$ within the block, and, thus, Bob will observe on average $(k-2)p(2|{\rm c})$ ``clicks'' from these signals.

Finally, with probability $p(2|{\rm c})$  the $k$-th signal of the block is $\ket{\varphi_2}$.  Since this signal does not include a vacuum pulse, she replaces it with $\ket{\varphi_{\rm vac}}$ (not shown in the figure). Then, Bob will observe $p_{\rm click}^{\rm decoy}(k-1|1)$ ``clicks'', because now Eve's block starts with $\ket{\varphi_1}$ and contains $k-2$ correctly identified signals $\ket{\varphi_j}$, with $j=0,\ldots,3$. 

Putting all together, we obtain the following recursive relation for $p_{\rm click}^{\rm decoy}(k|1)$,
\begin{eqnarray}\label{messi1b}
p_{\rm click}^{\rm decoy}(k|1)&=&\Big[p(0|{\rm c})(k-2)+p(1|{\rm c})(k-2)\nonumber \\
&+&p_{\rm click}^{\rm decoy}(k-1|1)\Big]p(2|{\rm c}).
\end{eqnarray}
By taking into account that $\sum_{i=0}^2p(i|{\rm c})=1$, $p(0|{\rm c})=p(1|{\rm c})$, and the fact that $p_{\rm click}^{\rm decoy}(1|1)=0$, from Eq.~(\ref{messi1b}) we find that 
\begin{eqnarray}\label{messi1}
p_{\rm click}^{\rm decoy}(k|1)&=&\frac{p(2|{\rm c})}{1-p(2|{\rm c})}\Bigg\{\Big[\big(1-p(2|{\rm c})\big)k+p(2|{\rm c})-2\Big] \nonumber \\
&+&p(2|{\rm c})^{k-1}\Bigg\}, 
\end{eqnarray}
for $k\geq{}2$.

\subsubsection{Parameter $p_{\rm click}^{\rm decoy}(k|0)$}

 \begin{figure}
\centering 
\centerline{\includegraphics*[scale=0.6]{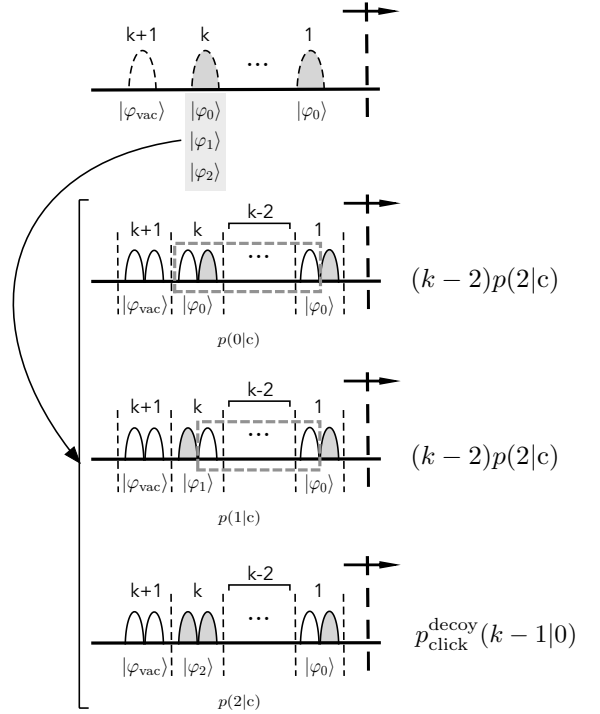}}
\caption{Illustration of the three sub-cases that we consider to evaluate $p_{\rm click}^{\rm decoy}(k|0)$. Eve replaces the first signal $\ket{\varphi_0}$ of the block with a vacuum signal $\ket{\varphi_{\rm vac}}$. With probability $p(0|{\rm c})$ the signal in the $k$-th position of the block is $\ket{\varphi_0}$. Then, Eve resends Bob the first $k$ signals, because now the block starts and ends with correctly identified vacuum pulses. Decoy signals can only be located in the positions $2$ to $k-1$ of the block. Therefore, the number of ``clicks'' at Bob's data line due to these signals is $(k-2)p(2|{\rm c})$. The other two sub-cases are analogous and we omit the details here for simplicity. In the figure, large ovals (drawn with a dashed line) represent signals, while small ovals (drawn with a solid line) represent optical pulses within a signal. The sub-blocks of signals that Eve resends to Bob are illustrated with grey dashed rectangles.
}
\label{fig:decoy2}
\end{figure}
This case is illustrated in Fig.~\ref{fig:decoy2}. The analysis is essentially equal to that of the previous section, and we omit the details here for simplicity. It can be shown that $p_{\rm click}^{\rm decoy}(k|0)$ satisfies
\begin{equation}\label{messi2}
p_{\rm click}^{\rm decoy}(k|0)=p_{\rm click}^{\rm decoy}(k|1).
\end{equation}

\subsubsection{Parameter $p_{\rm click}^{\rm decoy}(k|2)$}

When the first signal of a block is $\ket{\varphi_2}$, Eve replaces this signal with $\ket{\varphi_{\rm vac}}$ because $\ket{\varphi_2}$ does not contain a vacuum pulse. Now the new block has $k-1$ correctly identified signals $\ket{\varphi_i}$, with $i=0,1,2$. This means that 
\begin{equation}\label{messi3}
p_{\rm click}^{\rm decoy}(k|2)=p_{\rm click}^{\rm decoy}(k-1).
\end{equation}

\subsection{Parameters $p_{\rm click}^{\rm decoy}(k)$}

By combining Eqs.~(\ref{pclickFirstSignalBB_B})-(\ref{messi1})-(\ref{messi2})-(\ref{messi3}), we obtain that the parameter $p_{\rm click}^{\rm decoy}(k)$ can be written as
\begin{eqnarray}\label{messi4}
p_{\rm click}^{\rm decoy}(k)&=&\Big[\big(1-p(2|{\rm c})\big)k+p(2|{\rm c})-2+p_{\rm click}^{\rm decoy}(k-1)\Big]\nonumber \\
&\times&p(2|{\rm c})+p(2|{\rm c})^k.
\end{eqnarray}
This recursive relation can be solved by taking into account that $p_{\rm click}^{\rm decoy}(1)=0$. In particular, we find that
\begin{eqnarray}\label{messi5}
p_{\rm click}^{\rm decoy}(k)&=&\frac{p(2|{\rm c})}{1-p(2|{\rm c})}\Bigg\{\big(1-p(2|{\rm c})\big)k-2 \nonumber \\
&+&\Big[\big(1-p(2|{\rm c})\big)k+2p(2|{\rm c})\Big]p(2|{\rm c})^{k-1}\Bigg\}, \ \ \ \ \ \ 
\end{eqnarray}
for $k\geq{}2$.

\subsection{Evaluation}\label{poi}

To evaluate the effectiveness of this countermeasure, we optimize numerically the probabilities $\gamma_k\in[0,1]$, with $k=2,\ldots,M_{\rm max}$, and the parameter $\zeta\in[f,1]$. $\gamma_k$ characterizes the probability that Eve resends Bob a block that contains $k$ correctly identified signals, while $\zeta$ refers to the probability to correctly identify a decoy signal. The conclusive probability $p_{{\rm c},\zeta}$ of Eve's USD measurement depends on the value of $\zeta$ (see Appendix~\ref{meas_fri}). In the simulations below, we set $M_{\rm max}=10$, and disregard the effect of the dark counts of Bob's detectors.

\begin{figure}
\centering 
\centerline{\includegraphics*[scale=0.44]{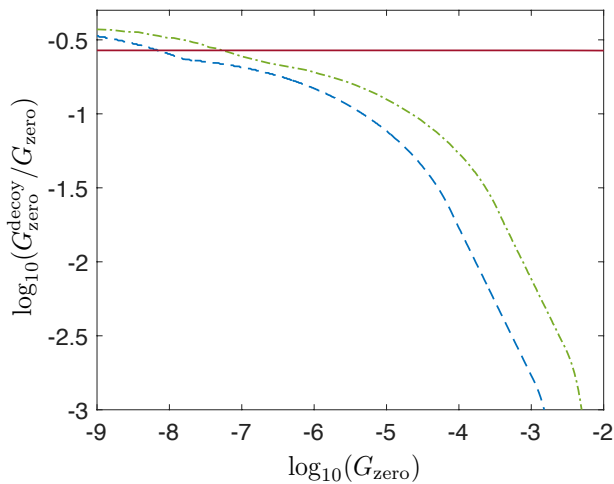}}
\caption{Maximum achievable value of $G_{\rm zero}^{\rm decoy}/G_{\rm zero}$ as a function of $G_{\rm zero}$ in logarithmic scale for the experimental parameters provided in Table~\ref{tab:korzhParams}. The dashed blue line (dash-dotted green line) corresponds to the case $\mu=0.06$ ($\mu=0.1$) shown in that table. The solid magenta line illustrates the expected detection rate of the decoy signals as a function of the gain in the absence of Eve for a typical channel model. In order for Eve to remain undetected, $G_{\rm zero}^{\rm decoy}/G_{\rm zero}$ must match the expected value given by the solid magenta line. This strongly reduces the maximum possible value of $G_{\rm zero}$ when compared to the results provided in \cite{zero_cow}. See the text for further details.}
\label{fig:t2bbcB}
\end{figure}

The results are illustrated in Fig.~\ref{fig:t2bbcB}. The dashed blue line (dash-dotted green line) shows the maximum value of $G_{\rm zero}^{\rm decoy}/G_{\rm zero}$ in logarithmic scale that is achievable by Eve, as a function of $G_{\rm zero}$ for the experimental parameters corresponding to $\mu=0.06$ ($\mu=0.1$) in Table~\ref{tab:korzhParams}. We note that the scenario considered in~\cite{zero_cow} corresponds to $\zeta=f$ (which provides $G_{\rm zero}^{\rm decoy}=0$). The solid magenta line shows the expected detection rate $G_{\rm decoy}/G$ of the decoy signals as a function of the expected gain $G=G_{\rm zero}$ for a typical channel model. In fact, in Fig.~\ref{fig:t2bbcB} there are two magenta lines, one for each of the two sets of experimental parameters provided in Table~\ref{tab:korzhParams}. However, these two lines cannot be distinguished with the resolution of the figure.

As shown in Fig.~\ref{fig:t2bbcB}, by monitoring the detection rates, Alice and Bob can dramatically decrease the value of $G_{\rm zero}$, and thus increase the maximum achievable distance $L_{\rm zero}$. This is so because now Eve must guarantee that $G_{\rm zero}^{\rm decoy}=G_{\rm decoy}$ for a certain $G=G_{\rm zero}$. To obtain $L_{\rm zero}$, we use the channel model described in Appendix~\ref{exp}. The values of $G_{\rm zero}$ (and the corresponding $L_{\rm zero}$) that satisfy this latter condition are provided in Table~\ref{tab:comparisonBc}. 
\begin{table}
\centering
\begin{tabular}{ |l|c|l|c| } 
 \hline
       & $\log_{10}(G_{\rm zero})$ & Att. [dB] & $L_{\rm zero}$(km)  \\ 
 \hline
$\mu=0.06$                  & -8.16  & $\ \ \approx 62.9$ & $\approx 387$   \\        
\hline
$\mu=0.1$                  & -7.27  & $\ \ \approx 57.1$ & $\approx 340$  \\ 
 \hline
\end{tabular}
\caption{Values of $G=G_{\rm zero}$ for which $G_{\rm zero}^{\rm decoy}=G_{\rm decoy}$. The meaning of the different parameters coincides with that provided in Table~\ref{tab:comparison}.}\label{tab:comparisonBc}
\end{table}
By comparing these results with those in Table~\ref{tab:comparison}, we find that $L_{\rm zero}$ is now more than eight times larger than the case where no detection rates are monitored, which is remarkable. 

\begin{figure}
\centering 
\centerline{\includegraphics*[scale=0.44]{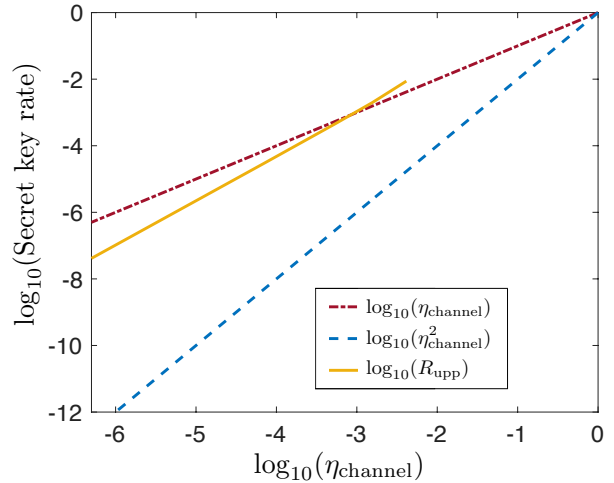}}
\caption{Upper bound $R_{\rm upp}$ on the secret key rate of COW-QKD as a function of the channel transmittance $\log_{10}(\eta_{\rm channel})$ when Alice and Bob monitor the detection rates of their signals. In the simulations we assume that $f=0.155$. For comparison, this figure includes as well the curves for linear and quadratic scaling in $\eta_{\rm channel}$.}
\label{fig:upperbound_decoy}
\end{figure}

To conclude this section, we evaluate the simple upper bound on the secret key rate of COW-QKD derived in~\cite{upper}. It reads
\begin{equation}\label{upp_bound_trivial}
K< (1-f)\eta_{\rm channel}\mu_{\rm max}(f)\equiv{}R_{\rm upp},
\end{equation}
where $\eta_{\rm channel}$ refers to the channel transmittance, and $\mu_{\rm max}(f)$ is the maximum allowed intensity for Alice's signals such that Eve's zero-error attack is not possible, {\it i.e.}, the expected gain in the absence of Eve is greater than $G_{\rm zero}$. We note that in Eq.~(\ref{upp_bound_trivial}) we implicitly assume the optimistic scenario in which the efficiency of Bob's detectors is $\eta_{\rm det}=1$. For further details, we refer the reader to~\cite{upper}. The results are illustrated in Fig.~\ref{fig:upperbound_decoy}. Remarkably, now the upper bound on the secret key rate scales close to $O(\eta^{4/3}_{\rm channel})$ with the channel transmittance (instead of quadratically, as is the case in the original COW scheme~\cite{upper}). 

\section{Four-state COW-QKD}\label{four}

In this section, we study a third possible countermeasure against zero-error attacks. The idea is to increase the number of signals sent by Alice to reduce the probability that Eve can unambiguously identify them. Precisely, we shall consider the situation in which Alice sends Bob an additional decoy signal $\ket{\varphi_3}=\ket{0}\ket{0}$~\cite{foot}. This solution has been evaluated in~\cite{cow_zer} against a restricted class of zero-error attacks. 

Alice now prepares her signals with probabilities 
\begin{eqnarray}
p_{\ket{\varphi_0}}&=&p_{\ket{\varphi_1}}=\frac{1-f}{2}, \nonumber \\
p_{\ket{\varphi_2}}&=&f_{\rm d}, \nonumber \\
p_{\ket{\varphi_3}}&=&f-f_{\rm d}\equiv f_{\rm v},
\end{eqnarray}
with $f,f_{\rm d}\in (0,1)$ and $f>f_{\rm d}$.  

To calculate the gain $G_{\rm zero}$ given by Eq.~(\ref{gzer}), we need to determine the parameters $p_{\rm c}$ and $p_{\rm click}(k)$. The conclusive probability $p_{\rm c}$ corresponding to the optimal USD measurement is obtained in Appendix~\ref{opt_usd}. Below, we calculate $p_{\rm click}(k)$.

\subsection{Probabilities $p_{\rm click}(k)$}

In this scenario, the probabilities $p(j|{\rm c})$ are given by
\begin{eqnarray}\label{cond_prob}
p(0|{\rm c})&=&p(1|{\rm c})=\frac{(1-f_{\rm d}-f_{\rm v})q_{\rm s}}{2p_{\rm c}}, \nonumber \\
p(2|{\rm c})&=&\frac{f_{\rm d} q^{\rm d}_{\rm s}}{p_{\rm c}}, \nonumber \\ 
p(3|{\rm c})&=&\frac{f_{\rm v} q^{\rm v}_{\rm s}}{p_{\rm c}},
\end{eqnarray}
where the terms $q_{\rm s}$, $q^{\rm d}_{\rm s}$, and $q^{\rm v}_{\rm s}$ are defined in Table~\ref{table1n} in Appendix~\ref{opt_usd}. They correspond to the probabilities that Eve's optimal USD measurement provides a conclusive result when measuring the key generation signals $\ket{\varphi_0}$ or $\ket{\varphi_1}$, and the decoy signals $\ket{\varphi_2}$ and $\ket{\varphi_3}$, respectively. 

To calculate $p_{\rm click}(k)$, we consider four different cases, depending on the result obtained by Eve for the first signal of a block that contains $k$ consecutive conclusive measurement results, and a vacuum signal $\ket{\varphi_{\rm vac}}=\ket{0}\ket{0}$ in its $k+1$th position. In particular, from Eq.~(\ref{pclickFirstSignalB}) we have that
\begin{eqnarray}\label{pclickFirstSignal}
p_{\rm click}(k)&=&\sum_{j=0}^3p(j|{\rm c})p_{\rm click}(k|j).
\end{eqnarray}
Below we compute $p_{\rm click}(k|j)$. Like in Sec.~\ref{det_rates}, we shall consider that Eve resends Bob coherent pulses of very high intensity such that he obtains a detection ``click'' with basically unit probability.

\subsubsection{Parameter $p_{\rm click}(k|1)$}

\begin{figure}
\centering 
\centerline{\includegraphics*[scale=0.6]{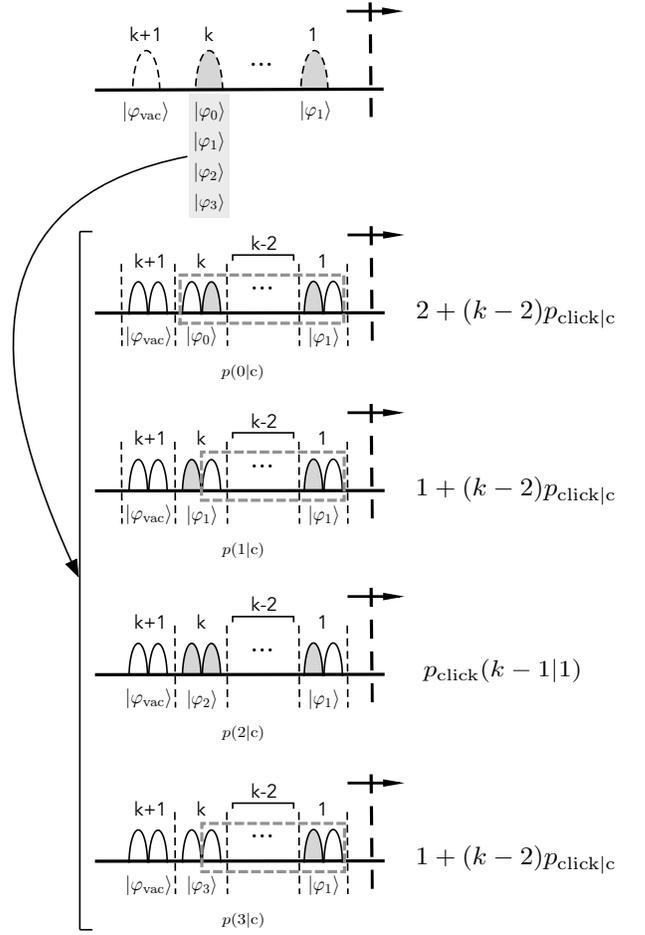}}
\caption{Illustration of the four sub-cases that we consider to evaluate $p_{\rm click}(k|1)$. With probability $p(0|{\rm c})$ the signal in the $k$-th position of the block is $\ket{\varphi_0}$. In this case, Eve resends Bob all the $k$ conclusive signals, as the block starts and ends with correctly identified vacuum pulses. The number of ``clicks'' at Bob's data line is then $2+(k-2)p_{\rm click|c}$, where $p_{\rm click|c}$ is defined in the text. The other three sub-cases are analogous and are described in the text. In the figure, large ovals (drawn with a dashed line) represent signals, while small ovals (drawn with a solid line) represent optical pulses within a signal. The sub-blocks of signals that Eve resends to Bob are illustrated with grey dashed rectangles.
}
\label{fig:firstSignal1a}
\end{figure}

This scenario is depicted in Fig.~\ref{fig:firstSignal1a}, which also shows the number of ``clicks'' that Bob observes in his data line for each of the four sub-cases considered in that figure. 
Since the first optical pulse of $\ket{\varphi_1}$ is a vacuum pulse, the longest sub-block that starts and ends with correctly identified vacuum {\it pulses} (if there is any) includes $\ket{\varphi_1}$. 

The probability that a signal correctly identified by Eve in positions $2$ to $k-1$ within the block produces a ``click'' at Bob's data line, which we shall denote by $p_{\rm click|c}$, is given by
\begin{equation}\label{eq_cli}
p_{\rm click|c}=\sum_{j=0}^3 p(j|{\rm c})p_{\rm click|j{\rm c}}=\sum_{j=0}^2 p(j|{\rm c})=1-p(3|{\rm c}),
\end{equation}
where $p_{\rm click|j{\rm c}}$ is the probability that the signal $\ket{\varphi_j}$ correctly identified by Eve results in a ``click'' at Bob's data line. In Eq.~(\ref{eq_cli}), we use $\sum_{j=0}^3 p(j|{\rm c})=1$ together with the fact that, as already mentioned, Eve's resent signals satisfy $p_{\rm click|j{\rm c}}=1$ for all $j=0,1,2$, and  $p_{\rm click|3{\rm c}}=0$. Here, note that double ``clicks'' in the data line within a signal duration are randomly assigned by Bob to single ``clicks''. 

With probability $p(0|{\rm c})$ the signal in the $k$-th position of the block is $\ket{\varphi_0}$. In this case, Eve resends Bob all the $k$ conclusive signals in the block, as it starts and ends with correctly identified vacuum pulses. Bob will observe two detection ``clicks'' from the signals located in the first and in the $k$-th position of the block, while for each of the other $k-2$ signals located in the middle between them, Bob will observe on average $p_{\rm click|c}$ ``clicks''. That is, the average number of ``clicks'' is $2+(k-2)p_{\rm click|c}$.

On the other hand, if the $k$-th signal of the block is $\ket{\varphi_1}$, which happens with probability $p(1|{\rm c})$, then Eve resends Bob the first $k-1$ conclusive signals in the block because such sub-block has correctly identified vacuum pulses on its edges. Also, she replaces the $k$-th signal with $\ket{\varphi_{\rm vac}}$ (not shown in Fig.~\ref{fig:firstSignal1a}). Therefore, Bob will observe one ``click'' from the first signal of the block, and $p_{\rm click|c}$ ``clicks'' for each of the $k-2$ signals in positions $2,\ldots,k-1$. That is, the average number of ``clicks'' is $1+(k-2)p_{\rm click|c}$. 

Likewise, with probability $p(2|{\rm c})$  the $k$-th signal of the block is $\ket{\varphi_2}$. Then, Eve replaces that signal with $\ket{\varphi_{\rm vac}}$  (not shown in Fig.~\ref{fig:firstSignal1a}) because it does not include a vacuum pulse. This means that Bob will observe $p_{\rm click}(k-1|1)$ ``clicks'', since now Eve's block starts with $\ket{\varphi_1}$ and includes $k-2$ correctly identified signals $\ket{\varphi_j}$, with $j=0,\ldots,3$. 

Finally, if the $k$-th signal of the block is $\ket{\varphi_3}$, which happens with probability $p(3|{\rm c})$, the situation is analogous to that in which that signal is $\ket{\varphi_1}$. That is, Eve resends Bob the first $k-1$ conclusive signals in the block because such sub-block has correctly identified vacuum pulses on its edges. Bob will observe on average $1+(k-2)p_{\rm click|c}$ ``clicks'' in this case. 

Putting all together, we find the following recursive relation for $p_{\rm click}(k|1)$,
\begin{eqnarray}\label{recursive_pc1}
p_{\rm click}(k|1)&=&p(0|{\rm c})[2+(k-2)p_{\rm click|c}]+p(1|{\rm c})\nonumber \\
&\times&[1+(k-2)p_{\rm click|c}]+p(2|{\rm c})p_{\rm click}(k-1|1)\nonumber \\
&+&p(3|{\rm c})[1+(k-2)p_{\rm click|c}].
\end{eqnarray} 

After some algebra, and taking into account that $p_{\rm click}(1|1)=0$, we obtain from Eq.~(\ref{recursive_pc1}) that
\begin{eqnarray}\label{final_pc1}
p_{\rm click}(k|1)&=&\frac{1}{p(2|{\rm c})\big[p(2|{\rm c})-1\big]}\bigg\{\left[1-p(0|{\rm c})-2p(2|{\rm c})\right]\nonumber \\
&\times&p(2|{\rm c})^k+p(2|{\rm c})\big[-1-p(0|{\rm c})+p(2|{\rm c}) \nonumber\\
&+&\left[2+(k-1)p(2|{\rm c})-k\right]p_{\rm click|c}\big]\bigg\}.
\end{eqnarray}

\subsubsection{Parameter $p_{\rm click}(k|0)$}

\begin{figure}
\centering 
\centerline{\includegraphics*[scale=0.6]{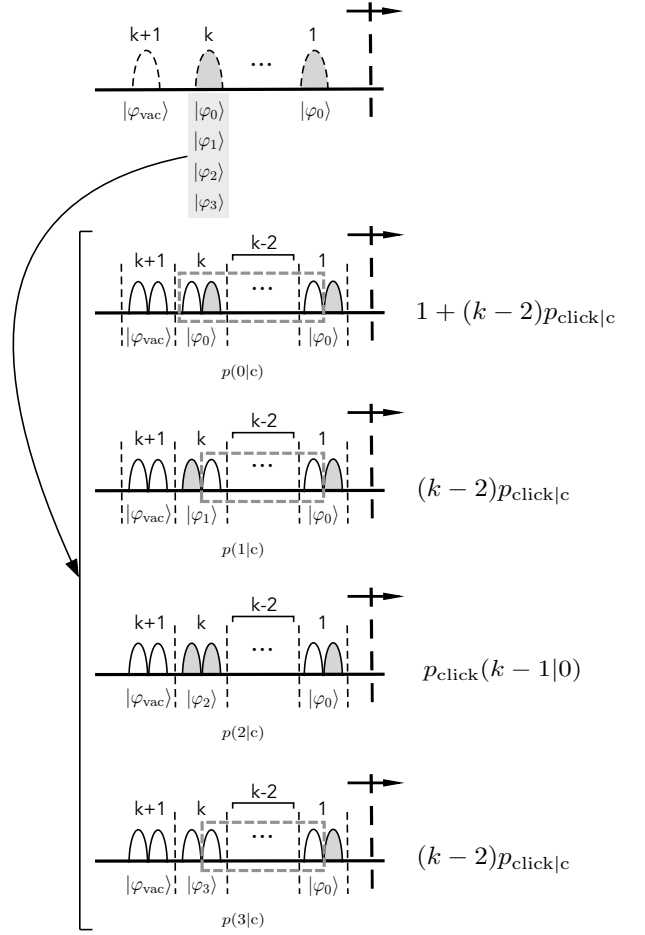}}
\caption{Illustration of the four sub-cases that we consider to evaluate $p_{\rm click}(k|0)$. With probability $p(0|{\rm c})$ the $k$-th signal of the block is $\ket{\varphi_0}$. In this case, Eve resends Bob the $k-1$ conclusive signals from position $2$ to position $k$, while the first signal is replaced with $\ket{\varphi_{\rm vac}}$ (not shown in the figure). The number of ``clicks'' at Bob's side is then $1+(k-2)p_{\rm click|c}$. The other three sub-cases are analogous and are described in the text. In the figure, large ovals (drawn with a dashed line) represent signals, while small ovals (drawn with a solid line) represent optical pulses within a signal. The sub-blocks of signals that Eve resends to Bob are illustrated with grey dashed rectangles.}
\label{fig:firstSignal1b}
\end{figure}

This case is very similar to the previous one, and is illustrated in Fig.~\ref{fig:firstSignal1b}. With probability $p(0|{\rm c})$, the $k$-th signal of the block is $\ket{\varphi_0}$. Then, Eve resends Bob the $k-1$ conclusive signals from positions $2$ to $k$, and she replaces the first signal with $\ket{\varphi_{\rm vac}}$ (not shown in the figure). Bob will observe one ``click'' from the $k$-th signal, and $p_{\rm click|c}$ ``clicks'' for each of the other $k-2$ signals. That is, the average number of ``clicks'' is $1+(k-2)p_{\rm click|c}$.

If the $k$-th signal of the block is $\ket{\varphi_1}$, which happens with probability $p(1|{\rm c})$, Eve resends Bob the $k-2$ conclusive signals from positions $2$ to $k-1$, and she replaces the first and the $k$th signals with $\ket{\varphi_{\rm vac}}$ (not shown in the figure). This means that the average number of ``clicks'' is $(k-2)p_{\rm click|c}$.

With probability $p(2|{\rm c})$, the $k$-th signal of the block is $\ket{\varphi_2}$. Then, Eve replaces that signal with $\ket{\varphi_{\rm vac}}$ (not shown in the figure) because it does not include a vacuum pulse. This means that Bob will observe $p_{\rm click}(k-1|0)$ ``clicks', since now Eve's block starts with $\ket{\varphi_0}$ and includes $(k-2)$ correctly identified signals $\ket{\varphi_j}$, with $j=0,\ldots,3$. 

Finally, if the $k$-th signal of the block is $\ket{\varphi_3}$, which happens with probability $p(3|{\rm c})$, the situation is analogous to that in which that signal is $\ket{\varphi_1}$. Eve replaces the first and the $k$th signals with $\ket{\varphi_{\rm vac}}$ (not shown in the figure), and resends Bob the $k-2$ conclusive signals from positions $2$ to $k-1$, as such sub-block has correctly identified vacuum pulses on its edges. This means that Bob will observe on average $(k-2)p_{\rm click|c}$ ``clicks''. 

Putting all together, we obtain the following recursive relation for $p_{\rm click}(k|0)$,
\begin{eqnarray}\label{recursive_pc1b}
p_{\rm click}(k|0)&=&p(0|{\rm c})[1+(k-2)p_{\rm click|c}]+p(1|{\rm c})(k-2)\nonumber \\
&\times&p_{\rm click|c}+p(2|{\rm c})p_{\rm click}(k-1|0)+p(3|{\rm c}) \nonumber \\
&\times& (k-2)p_{\rm click|c}.
\end{eqnarray} 

Since $p_{\rm click}(1|0)=0$, from Eq.~(\ref{recursive_pc1b}) it can be shown that $p_{\rm click}(k|0)$ satisfies
\begin{eqnarray}\label{final_pc1bb}
p_{\rm click}(k|0)&=&\frac{1}{p(2|{\rm c})\big[p(2|{\rm c})-1\big]}\bigg\{\left[p(0|{\rm c})-p_{\rm click|c}\right]\nonumber \\
&\times&p(2|{\rm c})^k-p(2|{\rm c})\big[p(0|{\rm c})+\big[(1-k)p(2|{\rm c})\nonumber \\
&+&k-2\big]p_{\rm click|c}\big]\bigg\}.
\end{eqnarray}

\subsubsection{Parameter $p_{\rm click}(k|2)$}

If the first signal of a block is $\ket{\varphi_2}$, Eve always replaces it with $\ket{\varphi_{\rm vac}}$, as $\ket{\varphi_2}$ does not contain a vacuum pulse. The remaining sub-block has now $k-1$ conclusive results, each of which can be a signal $\ket{\varphi_j}$ with $j=0,\ldots,3$. That is, $p_{\rm click}(k|2)$ is given by the average number of ``clicks'' of such sub-block, 
\begin{equation}\label{final_pc2}
p_{\rm click}(k|2)=p_{\rm click}(k-1).
\end{equation}

\subsubsection{Parameter $p_{\rm click}(k|3)$}

This case is essentially equal to that of $p_{\rm click}(k|0)$, because when a block starts with a signal $\ket{\varphi_0}$, Eve always replaces that signal with $\ket{\varphi_{\rm vac}}$. We find, therefore, that 
\begin{equation}\label{mon}
p_{\rm click}(k|3)=p_{\rm click}(k|0).
\end{equation} 

\subsubsection{Parameter $p_{\rm click}(k)$}

\begin{figure*}[!t]\center
\resizebox{14cm}{!}{\includegraphics{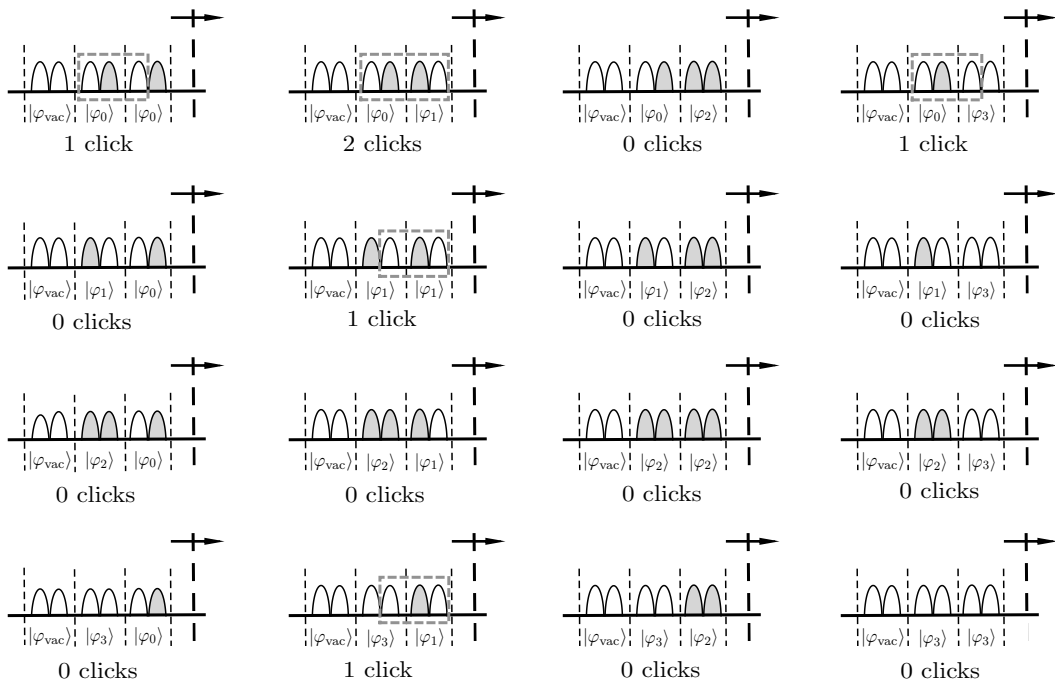}} \caption{Illustration of the sixteen possible cases for a block with $k=2$ consecutive conclusive measurement results, together with the number of ``clicks'' that Bob will observe in his data line. For example, in the first case, with probability $p(0|{\rm c})p(0|{\rm c})$ the signals in the block are $\ket{\varphi_{\rm vac}}\ket{\varphi_0}\ket{\varphi_0}$. This means that Eve can extract a sub-block surrounded by vacuum pulses by simply replacing the first signal of the block with $\ket{\varphi_{\rm vac}}$ (not shown in the figure). The number of ``clicks'' at Bob's data line is then one. The other cases are analogous. The sub-blocks of signals that Eve resends to Bob are illustrated with grey dashed rectangles.\label{fig:pclick2}}
\end{figure*}

By combining Eqs.~(\ref{pclickFirstSignal})-(\ref{final_pc1})-(\ref{final_pc1bb})-(\ref{final_pc2})-(\ref{mon}), we obtain the following recursive relation for $p_{\rm click}(k)$,
\begin{eqnarray}\label{final_pClick}
p_{\rm click}(k)&=&2p(0|{\rm c})\left[1-p(2|{\rm c})^{k-1}\right]+\big[k-2+(1-k)\nonumber \\
&\times&p(2|{\rm c})+p(2|{\rm c})^{k-1}\big]p_{\rm click|c}\nonumber \\
&+&p(2|{\rm c})p_{\rm click}(k-1).
\end{eqnarray}
To solve this equation for any $k>2$, we need to calculate the starting point of the recursion, $p_{\rm click}(2)$. For this, we consider the sixteen cases depicted in Fig.~\ref{fig:pclick2} with their {\it a priori} probabilities. We obtain
\begin{eqnarray}\label{pclick2}
p_{\rm click}(2)&=&p(0|{\rm c})^2+2p(0|{\rm c})p(1|{\rm c})+p(1|{\rm c})^2\nonumber \\
&+&p(3|{\rm c})p(1|{\rm c})+p(0|{\rm c})p(3|{\rm c}) \nonumber \\
&=&2p(0|{\rm c})\left[1-p(2|{\rm c})\right],
\end{eqnarray}
where in the second equality we use $p(0|{\rm c})=p(1|{\rm c})$ and $\sum_{j=0}^3p(j|{\rm c})=1$. 

By combining Eqs.~(\ref{final_pClick})-(\ref{pclick2}), we find that $p_{\rm click}(k)$ satisfies
\begin{widetext}
\begin{eqnarray}\label{FINAL_pClick}
p_{\rm click}(k)&=&\frac{1}{p(2|{\rm c})\left[p(2|{\rm c})-1\right]}\Big\{-2p(0|{\rm c})p(2|{\rm c})+p(2|{\rm c})^k\Big[2p(2|{\rm c})\Big(p(0|{\rm c})-p_{\rm click|c}\Big)
-k\Big(p(2|{\rm c})-1\Big)\Big(2p(0|{\rm c})-p_{\rm click|c}\Big)\Big]\nonumber \\
&+&\Big[2+k\Big(p(2|{\rm c})-1\Big)\Big]p(2|{\rm c})p_{\rm click|c}
\Big\}.
\end{eqnarray}
\end{widetext}
for any $k\geq 2$.

\subsection{Evaluation}

To evaluate the effectiveness of this countermeasure against zero-error attacks, we compute $G_{\rm zero}$ for the experimental parameters provided in Table~\ref{tab:korzhParams}. For this, we use Eqs.~(\ref{gzer})-(\ref{FINAL_pClick}), together with the value of $p_{\rm c}$ that corresponds to the optimal USD measurement calculated in Appendix~\ref{opt_usd}. Like in the previous sections, we fix $M_{\rm max}=10$ and, for simplicity, we disregard the effect of the dark counts of Bob's detectors. The results are illustrated in Table~\ref{tab:comparisonB}. They indicate that the resulting $G_{\rm zero}$ for the four-state COW-QKD protocol is significantly much smaller than that reported in~\cite{zero_cow} for the original scheme (see Table~\ref{tab:comparison}). 
\begin{table}
\centering
\begin{tabular}{ |l|c|l|c| } 
 \hline
       & $\log_{10}(G_{\rm zero})$ & Att. [dB] & $L_{\rm zero}$(km)  \\ 
 \hline
 $\mu=0.06$                  & -5.66  & $\ \ \approx 37.5$ & $\approx 231$   \\        
  \hline
 $\mu=0.1$                  & -4.79  & $\ \ \approx 31.9$ & $\approx 190$  \\ 
 \hline
\end{tabular}
\caption{$G_{\rm zero}$ for the experimental parameters provided in Table~\ref{tab:korzhParams}. For illustration purposes, we replace $f=0.155$ with $f_{\rm d}=0.1$ and $f_{\rm v}=0.055$. The parameter ``Att.'' ($L_{\rm zero}$) refers to the channel loss (distance) associated to $G_{\rm zero}$ if one considers the channel model described in Appendix~\ref{exp}.}\label{tab:comparisonB}
\end{table}

Table~\ref{tab:comparisonB} includes the maximum tolerable channel loss, and the maximum achievable distance $L_{\rm zero}$. To obtain $L_{\rm zero}$, we use the channel model described in Appendix~\ref{exp}, but now adapted to the four-state protocol. It can be shown that the expected value of the gain $G$ at Bob's data line in the absence of Eve's attack is now given by
\begin{equation}\label{gainExp_SM}
G=(1-f_{\rm d}-f_{\rm v})\left[1-e^{-\mu t_{\rm B}\eta_{\rm sys}}\right]+f_{\rm d}\left[1-e^{-2\mu t_{\rm B} \eta_{\rm sys}}\right],
\end{equation} 
where $\eta_{\rm sys}$ is the overall system's transmittance (see Appendix~\ref{exp}). The parameter $L_{\rm zero}$ corresponds to the distance for which $G=G_{\rm zero}$. 

\begin{figure}
\centering 
\centerline{\includegraphics*[scale=0.44]{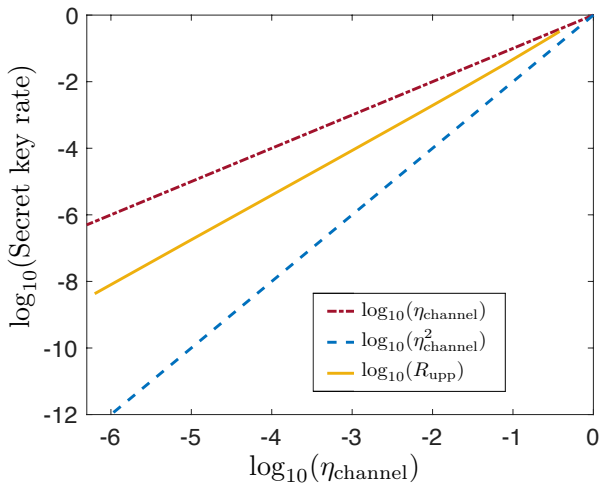}}
\caption{Upper bound $R_{\rm upp}$ on the secret key rate of the four-state COW-QKD protocol as a function of channel transmittance $\log_{10}(\eta_{\rm channel})$ when $f_{\rm d}=0.1$ and $f_{\rm v}=0.055$. For comparison, this figure includes as well the curves for linear and quadratic scaling in $\eta_{\rm channel}$.}
\label{fig:upperbound}
\end{figure}

Fig.~\ref{fig:upperbound} shows the upper bound on the secret key rate of COW-QKD derived in~\cite{upper}. In this scenario it has the form
\begin{equation}\label{upp_bound_trivial}
K< (1-f_{\rm d}-f_{\rm vac})\eta_{\rm channel}\mu_{\rm max}(f)\equiv{}R_{\rm upp}.
\end{equation}
The meaning of the different parameters has been given in Sec.~\ref{poi}. Again, like in Sec.~\ref{det_rates}, we obtain that now this upper bound scales close to $O(\eta^{4/3}_{\rm channel})$ with the channel transmittance, which contrasts with the quadratic scaling reported in~\cite{upper} for the original COW scheme. 

\section{Conclusion}\label{conclu}

In this paper, we have evaluated the effectiveness of three possible countermeasures for COW-QKD to foil zero-error attacks, in which the eavesdropper measures out all the emitted signals, and we have derived asymptotic upper security bounds for them. 

The first two countermeasures require that Alice and Bob monitor additional detection statistics. Precisely, we have considered the cases where they monitor the number of coincidences, and the detection rates of Alice's signals at Bob's data line. We have shown that the former countermeasure might allow them to approximately double the maximum achievable distance when compared to the original COW-QKD protocol, while the latter countermeasure improves the resulting performance even further, as now the upper security bounds on the secret key rate scale close to linear with the channel transmittance. This strongly contrasts with the quadratic scaling provided for the original scheme. Also, we have shown that a similar improvement could be obtained by increasing the number of transmitted signals, by adding, for instance, say a decoy vacuum signal.   

These countermeasures might be used to boost the performance of COW-QKD, though, for this, it would be necessary to derive lower security bounds that can confirm their merits. 

\section{Acknowledgements}

The authors wish to thank the company ID Quantique for very useful discussions, and for suggesting the consideration of this project. This work was funded by ID Quantique, the Galician Regional Government (consolidation of Research Units: AtlantTIC), the Spanish Ministry of Economy and Competitiveness (MINECO), and the Fondo Europeo de Desarrollo Regional (FEDER) through Grant No.~PID2020-118178RB-C21. 

\appendix

\section{Parameters $N_{ij}^{\rm ind}(k)$ and $N_{ij}^{\rm double}(k)$}\label{apA}

In this Appendix, we obtain the parameters $N_{ij}^{\rm ind}(k)$ and $N_{ij}^{\rm double}(k)$, with $i,j=0,1$, provided in Table~\ref{table1nnn}.

\begin{figure}
\centering 
\centerline{\includegraphics*[scale=0.8]{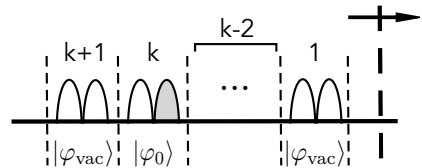}}
\caption{Illustration of a block of signals of the type ``00''. Eve replaces the signal $\ket{\varphi_0}$ located in the first position of the block with a signal $\ket{\varphi_{\rm vac}}$ because the first optical pulse of $\ket{\varphi_0}$ is not vacuum.}
\label{fig:param_N}
\end{figure}

Let us start with $N_{00}^{\rm ind}(k)$ and $N_{00}^{\rm double}(k)$. They represent, respectively, the average number of {\it individual} and {\it double} non-vacuum optical pulses contained in a sub-block of signals of the type ``00'' that Eve obtains from a block with $k$ correctly identified signals. This scenario is illustrated in Fig.~\ref{fig:param_N}.

As shown in the figure, when $k=2$, there is no double non-vacuum optical pulse ({\it i.e.}, $N_{00}^{\rm double}(2)=0$) and there is one individual non-vacuum optical pulse ({\it i.e.}, $N_{00}^{\rm ind}(2)=1$). Let us now consider the case $k\geq{}3$. A double non-vacuum optical pulse ({\it i.e.}, the combination $\ket{\varphi_0}\ket{\varphi_1}$) occurs with probability $p(0|{\rm c})p(1|{\rm c})=1/4$ {\it within} the block ({\it i.e.}, from positions $2$ to $k-1$, given that $k>3$). Since this combination takes two signals, say its first signal $\ket{\varphi_0}$ could only be located in $k-3$ different positions ({\it i.e.}, the positions $3$ to $k-1$) within the block. That is, on average we have $(k-3)/4$ double non-vacuum optical pulses within the block. If $k=3$, there is no double non-vacuum optical pulse within the block. Finally, let us consider the edges of the block. There cannot be a double non-vacuum optical pulse situated in the first and second position of the block because in the first position there is a vacuum signal $\ket{\varphi_{\rm vac}}$. On the other hand, a double non-vacuum optical pulse could be located in positions $k$ and $k-1$. Since in position $k$ there is $\ket{\varphi_0}$, this happens with probability $p(1|{\rm c})=1/2$ ({\it i.e.}, the probability to have $\ket{\varphi_1}$ in position $k-1$). Putting all together, we obtain
\begin{equation}
N_{00}^{\rm double}(k)=\frac{k-3}{4}+\frac{1}{2}=\frac{k-1}{4},
\end{equation}
when $k\geq{}3$.

A block of the type ``00'' has $k-1$ signals that contain non-vacuum optical pulses ({\it i.e.}, the signals in positions $2$ to $k$ of the block, see Fig.~\ref{fig:param_N}). Since the signals that are not double are individual, we find, therefore, that
\begin{equation}\label{vac1}
N_{00}^{\rm ind}(k)=(k-1)-2N_{00}^{\rm double}(k)=\frac{k-1}{2},
\end{equation}
when $k\geq{}3$. The factor two that multiplies the term $N_{00}^{\rm double}(k)$ in Eq.~(\ref{vac1}) is due to the fact that a double non-vacuum optical pulse takes two signals. 

The other parameters $N_{ij}^{\rm ind}(k)$ and $N_{ij}^{\rm double}(k)$ in Table~\ref{table1nnn} can be calculated similarly, and we omit the details here for simplicity. 

\section{Calculation of $P_{\rm click}^{\rm ind}$ and $P_{\rm coin}^{\rm ind}$}\label{qwe_ap}

This corresponds to the scenario where Eve sends Bob an {\it individual} non-vacuum optical pulse in a state $\ket{\Phi^{\rm ind}}$ given by Eq.~(\ref{f1n}) surrounded by vacuum pulses.  

\begin{figure}
\centering 
\centerline{\includegraphics*[scale=0.50]{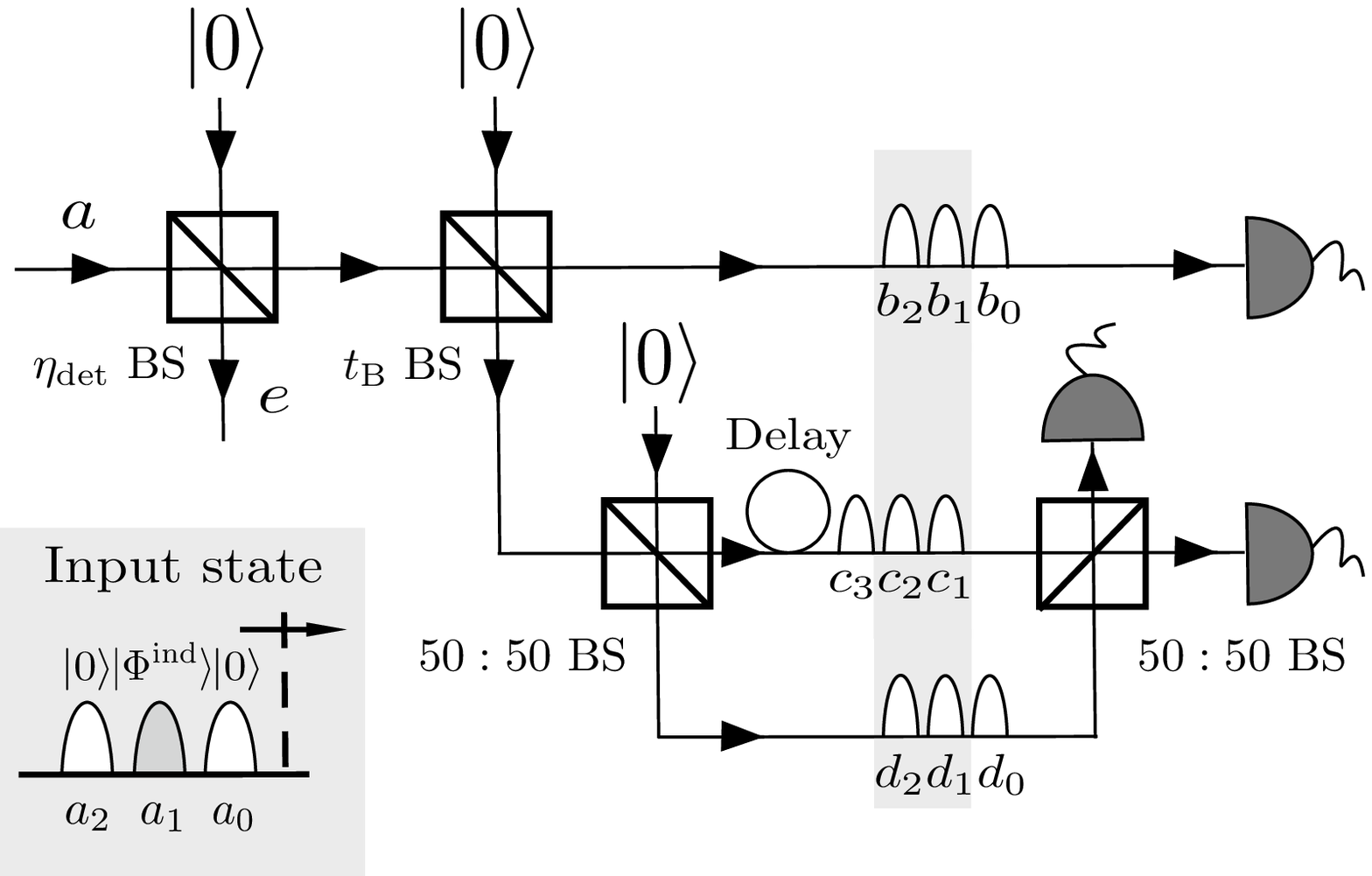}}
\caption{Schematic representation of the linear optics circuit that we use to calculate the probabilities $P_{\rm click}^{\rm ind}$ and $P_{\rm coin}^{\rm ind}$. The effect of the finite detection efficiency of Bob's detectors is modelled with a beamsplitter (BS) of transmittance $\eta_{\rm det}$ placed right before Bob's receiver. In doing so, we can now assume that all of his detectors have perfect detection efficiency. This fictitious beamsplitter cannot be controlled by Eve. We further assume that Eve sends Bob a non-vacuum optical pulse in the state $\ket{\Phi^{\rm ind}}$ in the mode $a_1$ (where the subindex ``1'' indicates the time instant), surrounded by vacuum pulses in the modes $a_0$ and $a_2$.  The non-vacuum optical pulse is illustrated in the figure with a grey oval. To determine $P_{\rm click}^{\rm ind}$ and $P_{\rm coin}^{\rm ind}$, we are interested in the state of the signal in the modes $b_2$, $b_1$, $c_2$, $c_1$, $d_2$ and $d_1$. In the figure, the states $\ket{0}$ represent vacuum pulses.
}
\label{fig:t4}
\end{figure}

For simplicity, in the calculations below, we assume that all detectors at Bob's side have the same detection efficiency $\eta_{\rm det}$. This means that we can model the effect of their finite detection efficiency by placing a beamsplitter of transmittance $\eta_{\rm det}$ right before Bob's receiver, and then we assume that all his detectors have now perfect detection efficiency. Since in Sec.~\ref{coincidences} we consider the trusted device scenario, this fictitious beamsplitter cannot be controlled by Eve. Moreover, we consider that Eve sends the signals to Bob with a transmitter located very close to his receiver, and thus they are not affected by channel loss. This situation is illustrated in Fig.~\ref{fig:t4}.

According to Eq.~(\ref{f1n}), the input state in modes $a_i$, where the subindex $i=0,1,2$, represents the time instant, is given by
\begin{equation}\label{f1}
\ket{\Psi}_{a_2a_1a_0}=\ket{0}_{a_2}\left(\sum_{n=0}^\infty q_n\ket{n}_{a_1}\right)\ket{0}_{a_0}.
\end{equation}
When this state enters the linear optics circuit illustrated in Fig.~\ref{fig:t4}, it can be shown that the output state at the optical modes $b_2$, $b_1$, $c_2$, $c_1$, $d_2$, $d_1$ and $e$ depicted in that figure can be written as 
\begin{equation}
\ket{\Phi}_{b_2b_1c_2c_1d_2d_1e}=\ket{\phi}_{b_1c_2d_1e}\ket{0}_{b_2c_1d_2},
\end{equation}
where the state $\ket{\phi}_{b_1c_2d_1e}$ has the form
\begin{eqnarray}
\ket{\phi}_{b_1c_2d_1e}&=&\sum_{n=0}^\infty \sum_{k=0}^n\sum_{l=0}^k\sum_{m=0}^l \binom{n}{k}\binom{k}{l}\binom{l}{m}\frac{q_n}{\sqrt{n!}}\nonumber \\
&\times&\sqrt{1-\eta_{\rm det}}^{n-k}\sqrt{\eta_{\rm det}}^k\sqrt{t_{\rm B}}^{k-l}\sqrt{\frac{1-t_{\rm B}}{2}}^l\nonumber \\
&\times&\sqrt{(n-k)!}\sqrt{(k-l)!}\sqrt{(l-m)!}\sqrt{m!}\nonumber \\
&\times&\ket{k-l}_{b_1}\ket{l-m}_{c_2}\ket{m}_{d_1}\ket{n-k}_e.
\end{eqnarray}
Here, $\ket{k-l}_{b_1}$ is the Fock state with $k-l$ photons in the mode $b_1$, and the other states are defined similarly. In the output modes $b_0$, $c_3$ and $d_0$ illustrated in Fig.~\ref{fig:t4} there are vacuum states. These states are not relevant for the discussion and calculations below, as we are only interested in the detection events that may occur in the time instances ``1'' and ``2''. 

To calculate $P_{\rm click}^{\rm ind}$ and $P_{\rm coin}^{\rm ind}$, we first trace out mode $e$ from the state $\ket{\phi}_{b_1c_2d_1e}\ket{0}_{b_2c_1d_2}$, and consider instead the state $\rho'_{b_2b_1c_2c_1d_2d_1}={\rm Tr}_e(\ket{\phi}\bra{\phi}_{b_1c_2d_1e})\otimes\ketbra{0}_{b_2c_1d_2}$. Moreover, since Bob's measurement operators on the modes $b_2$, $b_1$, $c_2$, $c_1$, $d_2$ and $d_1$ are diagonal in the Fock basis (due to the use of single-photon detectors), we can further simplify the calculations by replacing $\rho'_{b_2b_1c_2c_1d_2d_1}$ with a state ${\tilde \rho}_{b_2b_1c_2c_1d_2d_1}$ that is diagonal in the Fock basis, and whose diagonal elements coincide with those of $\rho'_{b_2b_1c_2c_1d_2d_1}$. That is, it follows that the measurement statistics provided by ${\tilde \rho}_{b_2b_1c_2c_1d_2d_1}$ match those of $\rho'_{b_2b_1c_2c_1d_2d_1}$ when both states are measured with Fock diagonal measurement operators. The state ${\tilde \rho}_{b_2b_1c_2c_1d_2d_1}$ has the form
\begin{equation}
{\tilde \rho}_{b_2b_1c_2c_1d_2d_1}={\tilde \rho}_{b_1c_2d_1}\otimes\ketbra{0}_{b_2c_1d_2},
\end{equation}
with ${\tilde \rho}_{b_1c_2d_1}$ having the form 
\begin{eqnarray}\label{Mon}
{\tilde \rho}_{b_1c_2d_1}&=&\sum_{n=0}^\infty \sum_{k=0}^n\sum_{l=0}^k\sum_{m=0}^l \binom{n}{k}\binom{k}{l}\binom{l}{m}\abs{q_n}^2\nonumber \\
&\times&(1-\eta_{\rm det})^{n-k}\eta_{\rm det}^kt_{\rm B}^{k-l}\left(\frac{1-t_{\rm B}}{2}\right)^l\nonumber \\
&\times&\ket{k-l}_{b_1}\bra{k-l}\otimes\ket{l-m}_{c_2}\bra{l-m} \nonumber \\
&\otimes&\ket{m}_{d_1}\bra{m}.
\end{eqnarray}

The state ${\tilde \rho}_{b_2b_1c_2c_1d_2d_1}$ can only produce a ``click'' in Bob's data line in the time instant ``1'' because in mode $b_2$ there is a vacuum state. This means that $P_{\rm click}^{\rm ind}$ can be expressed as
\begin{eqnarray}\label{resb}
P_{\rm click}^{\rm ind}&=&1-\bra{0}{\rm Tr}_{\it c_{\rm 2}d_{\rm 1}}\left({\tilde \rho}_{\it b_{\rm 1}c_{\rm 2}d_{\rm 1}}\right)\ket{0}_{{\it b}_1}.
\end{eqnarray}
By using Eqs.~(\ref{Mon})-(\ref{resb}) we obtain Eq.~(\ref{resA}).

For the same reason, a coincidence detection event is only possible in the time instant ``1''. Since there is vacuum in mode $c_1$, the probability that at least one of the two detectors $D_{\rm M1}$ and $D_{\rm M2}$ ``click'' at Bob's monitoring line, is equal to the probability to have a ``click'' in $d_1$ if we would had measured this mode directly. This means that $P_{\rm coin}^{\rm ind}$ can be written as
\begin{equation}\label{Mon2}
P_{\rm coin}^{\rm ind}={\rm Tr}\left({\tilde \rho}_{b_1c_2d_1}M_{b_1c_2d_1}\right),
\end{equation}
where the operator $M_{b_1c_2d_1}$ has the form 
\begin{equation}\label{asd}
M_{b_1c_2d_1}=(\openone_{b_1}-\ket{0}\bra{0}_{b_1})\otimes\openone_{c_2}\otimes(\openone_{d_1}-\ket{0}\bra{0})_{d_1},
\end{equation}
with $\openone$ denoting the identity operator. That is, $M_{b_1c_2d_1}$ corresponds to a simultaneous detection ``click'' in the modes $b_1$ and $d_1$. By using Eqs.~(\ref{Mon})-(\ref{Mon2})-(\ref{asd}), we obtain Eq.~(\ref{resA}).

\begin{figure}
\centering 
\centerline{\includegraphics*[scale=0.5]{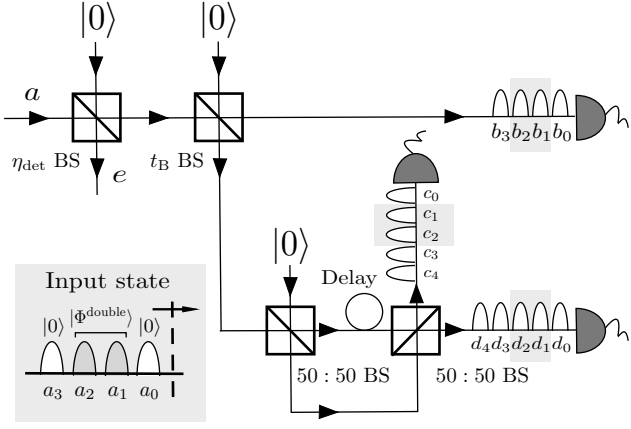}}
\caption{To calculate the probabilities $P_{\rm click}^{\rm double}$ and $P_{\rm coin}^{\rm double}$, we use the same linear optics circuit employed in Fig.~\ref{fig:t4}, which we reproduce in this figure as well. However, now Eve sends Bob two adjacent non-vacuum optical pulses in a joint state $\ket{\Phi^{\rm double}}$, surrounded by vacuum pulses. The two non-vacuum optical pulses are illustrated in the figure with two grey ovals. To determine $P_{\rm click}^{\rm double}$ and $P_{\rm coin}^{\rm double}$, we are interested in the state of the optical modes $b_2$, $b_1$, $c_2$, $c_1$, $d_2$ and $d_1$. In the figure, the states $\ket{0}$ represent vacuum pulses.
}
\label{fig:t5}
\end{figure}
\section{Calculation of $P_{\rm click}^{\rm double}$ and $P_{\rm coin}^{\rm double}$}\label{qwe_ap2}

This corresponds to the scenario where Eve sends Bob a {\it double} non-vacuum optical pulse in a state $\ket{\Phi^{\rm double}}$ given by Eq.~(\ref{Mon3}).  

Like in Appendix~\ref{qwe_ap}, we consider that Bob's detectors have detection efficiency $\eta_{\rm det}$, and Eve's signals are not affected by channel loss as she could send them to Bob with a transmitter located very close to him. This situation is illustrated in Fig.~\ref{fig:t5}.

According to Eq.~(\ref{Mon3}), the input state in modes $a_i$, where the subindex $i=0,1,2,3$, represents the time instant, is given by
\begin{equation}\label{f1bb}
\ket{\Psi}_{a_3a_2a_1a_0}=\ket{0}_{a_3}\left[\sum_{n=0}^\infty \frac{p_n}{\sqrt{n!}}\left(\frac{a_1^\dagger+a_2^\dagger}{\sqrt{2}}\right)^n\ket{0}_{a_{1}a_{2}}\right]\ket{0}_{a_0},
\end{equation}
where $a_1^\dagger$ ($a_2^\dagger$) denotes the creation operator associated to photons in mode $a$ at the time instant ``1'' (``2''). 

When this state enters the linear optics circuit illustrated in Fig.~\ref{fig:t5}, it can be shown that the output state 
at the optical modes $b_2$, $b_1$, $c_3$, $c_2$, $c_1$, $d_3$, $d_2$, $d_1$ and $e$ depicted in that figure is of the form 
\begin{equation}
\ket{\phi}_{b_2b_1c_3c_2c_1d_3d_2d_1e}=\ket{\phi}_{b_2b_1c_3c_1d_3d_2d_1e}\otimes\ket{0}_{c_2}, 
\end{equation}
where the state $\ket{\phi}_{b_2b_1c_3c_1d_3d_2d_1e}$ is given by
\begin{equation}\label{Mon5}
\ket{\phi}_{b_2b_1c_3c_1d_3d_2d_1e}=\sum_{n=0}^\infty p_n\ket{\tilde n}_{b_2b_1c_3c_1d_3d_2d_1e}, 
\end{equation}
and the state $\ket{\tilde n}_{b_2b_1c_3c_1d_3d_2d_1e}$, which has in total $n$ photons, has the form
\begin{widetext}
\begin{eqnarray}
\ket{\tilde n}_{b_2b_1c_3c_1d_3d_2d_1e}&=&\frac{1}{\sqrt{n!}\sqrt{2}^n}\left[\sqrt{2(1-\eta_{\rm det})}e^\dagger+\sqrt{\eta_{\rm det}t_{\rm B}}\left(b_2^\dagger+b_1^\dagger\right)+\frac{\sqrt{\eta_{\rm det}(1-t_{\rm B})}}{2}\left(-c_3^\dagger+c_1^\dagger+d_3^\dagger+2d_2^\dagger+d_1^\dagger\right)\right]^n \nonumber \\
&\times&\ket{0}_{b_2b_1c_3c_1d_3d_2d_1e}. 
\end{eqnarray}
\end{widetext}
In this equation, we have defined the creation operator $e^\dagger=(e_1^\dagger+e_2^\dagger)/\sqrt{2}$. 

In the output modes $b_3$, $b_0$, $c_4$, $c_0$, $d_4$ and $d_0$ illustrated in Fig.~\ref{fig:t5} there are vacuum states that are not relevant for the calculations below, as we are only interested in the detection events that may occur in the time instances ``1'' and ``2''. Also, we note that the fact that there is vacuum in mode $c_2$ confirms that the state $\ket{\Psi}_{a_3a_2a_1a_0}$ sent by Eve preserves the mode of COW-QKD.

As discussed in Appendix~\ref{qwe_ap}, Bob's measurement operators are diagonal in the Fock basis. This means that they cannot distinguish the state $\ket{\phi}_{b_2b_1c_3c_1d_3d_2d_1e}$ given by Eq.~(\ref{Mon5}) from a state $\rho_{b_2b_1c_3c_1d_3d_2d_1e}$ given by
\begin{equation}
\rho_{b_2b_1c_3c_1d_3d_2d_1e}=\sum_{n=0}^\infty \abs{p_n}^2 \ket{\tilde n}\bra{\tilde n}_{b_2b_1c_3c_1d_3d_2d_1e}.
\end{equation}

After tracing out the systems $c_3$, $d_3$ and $e$ from $\rho_{b_2b_1c_3c_1d_3d_2d_1e}\otimes\ket{0}\bra{0}_{c_2}$, we obtain
\begin{eqnarray}\label{vasallo}
\rho_{b_2b_1c_2c_1d_2d_1}&=&{\rm Tr}_{d_3c_3e}\left(\rho_{b_2b_1c_3c_1d_3d_2d_1e}\otimes\ket{0}\bra{0}_{c_2}\right)\nonumber \\
&=&\sum_{n=0}^\infty \frac{\abs{p_n}^2}{4^n}\sum_{k=0}^n\binom{n}{k}\left[4-(3+t_{\rm B})\eta_{\rm det}\right]^{n-k}\nonumber \\
&\times&\left[(3+t_{\rm B})\eta_{\rm det}\right]^k\ket{k}\bra{k}_{b_2b_1c_1d_2d_1}\nonumber \\
&\otimes&\ket{0}\bra{0}_{c_2}, 
\end{eqnarray}
where the state $\ket{k}_{b_2b_1c_1d_2d_1}$, which has $k$ photons in total, has the form
\begin{widetext}
\begin{eqnarray}
\ket{k}_{b_2b_1c_1d_2d_1}&=&\frac{\sqrt{2}^k}{\sqrt{k!}\sqrt{3+t_{\rm B}}^k}\sum_{l=0}^k\sum_{m=0}^l\sum_{o=0}^{k-l}\sum_{p=0}^{l-m} \binom{k}{l}\binom{l}{m}\binom{k-l}{o}\binom{l-m}{p} \frac{\sqrt{t_{\rm B}}^{m+o}\sqrt{1-t_{\rm B}}^{k-m-o}}{2^{l-m}}\sqrt{m!}\sqrt{p!}\nonumber \\
&\times&\sqrt{(l-m-p)!}\sqrt{o!}\sqrt{(k-l-o)!}\ket{o}_{b_2}\ket{m}_{b_1}\ket{p}_{c_1}\ket{k-l-o}_{d_2}\ket{l-m-p}_{d_1}.
\end{eqnarray}
\end{widetext}

This state has still non-zero off-diagonal terms in the Fock basis. So, to simplify the calculations below, we use again the fact that Bob's measurement operators are diagonal in the Fock basis. This means that we can consider instead a state $\ket{\tilde k}\bra{\tilde k}_{b_2b_1c_1d_2d_1}$ that is diagonal in the Fock basis, and whose diagonal elements coincide with those of $\ket{k}\bra{k}_{b_2b_1c_1d_2d_1}$. This state is given by
\begin{widetext}
\begin{eqnarray}
\ket{\tilde k}\bra{\tilde k}_{b_2b_1c_1d_2d_1}&=&\frac{2^k}{k!(3+t_{\rm B})^k}\sum_{l=0}^k\sum_{m=0}^l\sum_{o=0}^{k-l}\sum_{p=0}^{l-m} \binom{k}{l}^2\binom{l}{m}^2\binom{k-l}{o}^2\binom{l-m}{p}^2 \frac{t_{\rm B}^{m+o}(1-t_{\rm B})^{k-m-o}}{4^{l-m}}m!p!(l-m-p)!\nonumber \\
&\times&o!(k-l-o)!\ket{o}\bra{o}_{b_2}\otimes\ket{m}\bra{m}_{b_1}\otimes\ket{p}\bra{p}_{c_1}\otimes\ket{k-l-o}\bra{k-l-o}_{d_2}\otimes\ket{l-m-p}\bra{l-m-p}_{d_1}. \nonumber \\
\end{eqnarray}
\end{widetext}
That is, it follows that the measurement statistics provided by $\ket{k}_{b_2b_1c_1d_2d_1}$ match those of $\ket{\tilde k}\bra{\tilde k}_{b_2b_1c_1d_2d_1}$ when both states are measured with Fock diagonal measurement operators.

Therefore, to obtain $P_{\rm click}^{\rm double}$ and $P_{\rm coin}^{\rm double}$, we use the following state
\begin{eqnarray}\label{Mon6}
{\tilde \rho}_{b_2b_1c_2c_1d_2d_1}&=&\sum_{n=0}^\infty \frac{\abs{p_n}^2}{4^n}\sum_{k=0}^n\binom{n}{k}\left[4-(3+t_{\rm B})\eta_{\rm det}\right]^{n-k}\nonumber \\
&\times&\left[(3+t_{\rm B})\eta_{\rm det}\right]^k\ket{\tilde k}\bra{\tilde k}_{b_2b_1c_1d_2d_1} \nonumber \\
&\otimes&\ket{0}\bra{0}_{c_2}.
\end{eqnarray}

The probability that Bob observes a single ``click'' in his data line can then be calculated as
\begin{equation}\label{Mon7}
P_{\rm single-click}={\rm Tr}\left({\tilde \rho}_{b_2b_1c_2c_1d_2d_1}M_{b_2b_1c_2c_1d_2d_1}^{\rm single-click}\right),
\end{equation}
where the operator $M_{b_2b_1c_2c_1d_2d_1}^{\rm single-click}$ is given by
\begin{eqnarray}\label{eq_ssa}
M_{b_2b_1c_2c_1d_2d_1}^{\rm single-click}&=&\left(\openone_{b_2}-\ket{0}\bra{0}_{b_2}\right)\otimes\ket{0}\bra{0}_{b_1}\otimes\openone_{c_2c_1d_2d_1}\nonumber \\
&+&\ket{0}\bra{0}_{b_2}\otimes\left(\openone_{b_1}-\ket{0}\bra{0}_{b_1}\right)\otimes\openone_{c_2c_1d_2d_1}. \nonumber \\
\end{eqnarray}
The first (second) term on the right hand side of Eq.~(\ref{eq_ssa}) corresponds to one detection ``click'' in $b_2$ ($b_1$) and no ``click'' in $b_1$ ($b_2$). By combining Eqs.~(\ref{Mon6})-(\ref{Mon7})-(\ref{eq_ssa}), we obtain
\begin{eqnarray}\label{zxc}
P_{\rm single-click}&=&2\sum_{n=0}^\infty \abs{p_n}^2\Bigg[\left(1-\frac{t_{\rm B}\eta_{\rm det}}{2}\right)^n\nonumber \\
&-&(1-t_{\rm B}\eta_{\rm det})^n\Bigg].
\end{eqnarray}

Likewise, we have that the probability that Bob observes two detection ``clicks'' in his data line, one in $b_1$ and one in $b_2$, has the form
\begin{equation}\label{Mon8}
P_{\rm double-click}={\rm Tr}\left({\tilde \rho}_{b_2b_1c_2c_1d_2d_1}M_{b_2b_1c_2c_1d_2d_1}^{\rm double-click}\right),
\end{equation}
where the operator $M_{b_2b_1c_2c_1d_2d_1}^{\rm double-click}$ is given by
\begin{eqnarray}\label{Mon9}
M_{b_2b_1c_2c_1d_2d_1}^{\rm double-click}&=&\left(\openone_{b_2}-\ket{0}\bra{0}_{b_2}\right)\otimes\left(\openone_{b_1}-\ket{0}\bra{0}_{b_1}\right)\nonumber \\
&\otimes&\openone_{c_2c_1d_2d_1}.
\end{eqnarray}
By combining Eqs.~(\ref{Mon6})-(\ref{Mon8})-(\ref{Mon9}), we obtain
\begin{eqnarray}\label{zxc1}
P_{\rm double-click}&=&1-\sum_{n=0}^\infty \abs{p_n}^2\Bigg[2\left(1-\frac{t_{\rm B}\eta_{\rm det}}{2}\right)^n\nonumber \\
&-&(1-t_{\rm B}\eta_{\rm det})^n\Bigg].
\end{eqnarray}

The average number of ``clicks'' in Bob's data line can be written as
\begin{eqnarray}\label{zxc2}
P_{\rm click}^{\rm double}&=&P_{\rm single-click}+2P_{\rm double-click}. 
\end{eqnarray}
The factor two that multiples $P_{\rm double-click}$ arises because the two non-vacuum optical pulses sent by Eve correspond to two different signals sent by Alice. That is, in this case Bob do not assign double ``clicks'' to single ``click'' events. By combining Eqs.~(\ref{zxc})-(\ref{zxc1})-(\ref{zxc2}), we obtain Eq.~(\ref{resB}). 

The parameter $P_{\rm coin}^{\rm double}$ can also be expressed as the sum of two terms
\begin{equation}\label{zam2}
P_{\rm coin}^{\rm double}=P_{\rm single-coin}+2P_{\rm double-coin},
\end{equation}
where $P_{\rm single-coin}$ ($P_{\rm double-coin}$) denotes the probability to observe a coincidence detection event only in one of the time instants ``1'' or ``2'' (in both time instants). 

The probability $P_{\rm single-coin}$ is given by
\begin{eqnarray}\label{dieg}
P_{\rm single-coin}&=&P_{\rm click}^{b_1,c_1d_1,{\bar b_2},\bar{c_2d_2}}+P_{\rm click}^{b_1,c_1d_1,b_2,\bar{c_2d_2}}\nonumber \\
&+&P_{\rm click}^{b_1,c_1d_1,{\bar b_2},c_2d_2}+P_{\rm click}^{{\bar b_1},{\bar{c_1d_1}},b_2,c_2d_2} \nonumber \\
&+&P_{\rm click}^{b_1,{\bar{c_1d_1}},b_2,c_2d_2}+P_{\rm click}^{{\bar b_1},c_1d_1,b_2,c_2d_2},\ \ \ \ \ 
\end{eqnarray}
where $P_{\rm click}^{b_1,c_1d_1,{\bar b_2},\bar{c_2d_2}}$ denotes the probability that Bob observes a ``click'' in the modes $b_1$, $c_1$ and/or $d_1$, and no ``click'' in the modes $b_2$, $c_2$ and $d_2$. That is, this is the probability that Bob observes a coincidence detection event in the time instant ``1'' and no ``click'' in the time instant ``2''. This probability can be expressed as
\begin{equation}\label{Mon10}
P_{\rm click}^{b_1,c_1d_1,{\bar b_2},\bar{c_2d_2}}={\rm Tr}\left({\tilde \rho}_{b_2b_1c_2c_1d_2d_1}M_{b_1,c_1d_1,{\bar b_2},\bar{c_2d_2}}\right),
\end{equation}
where the state ${\tilde \rho}_{b_2b_1c_2c_1d_2d_1}$ is given by Eq.~(\ref{Mon6}), and the operator $M_{b_1,c_1d_1,{\bar b_2},\bar{c_2d_2}}$ has the form
\begin{eqnarray}\label{Mon11}
M_{b_1,c_1d_1,{\bar b_2},\bar{c_2d_2}}&=&\ket{0}\bra{0}_{b_2}\otimes\left(\openone_{b_1}-\ket{0}\bra{0}_{b_1}\right)\nonumber \\
&\otimes&\left(\openone_{c_1d_1}-\ket{00}\bra{00}_{c_1d_1}\right)\otimes\ket{0}\bra{0}_{d_2}\nonumber \\
&\otimes&\ket{0}\bra{0}_{c_2}.
\end{eqnarray}
By combining Eqs.~(\ref{Mon6})-(\ref{Mon10})-(\ref{Mon11}), we obtain
\begin{widetext}
\begin{equation}\label{zam}
P_{\rm click}^{b_1,c_1d_1,{\bar b_2},\bar{c_2d_2}}=\sum_{n=0}^\infty \abs{p_n}^2\left\{\left[1-\frac{\eta_{\rm det}}{2}\right]^n-\left[1-\frac{(3-t_{\rm B})\eta_{\rm det}}{4}\right]^n-\left[1-\frac{(1+t_{\rm B})\eta_{\rm det}}{2}\right]^n+\left[1-\frac{(3+t_{\rm B})\eta_{\rm det}}{4}\right]^n\right\}.
\end{equation}
\end{widetext}

The other probabilities that appear in Eq.~(\ref{dieg}) are defined similarly, and can be obtained by following an analogous procedure to that used to calculate $P_{\rm click}^{b_1,c_1d_1,{\bar b_2},\bar{c_2d_2}}$. For simplicity, we omit the details here. The results are shown in Table~\ref{table1nnnBB}. 
\begin{table}[h!]
  \centering
  \begin{tabular}{l}
    \hline\hline
$P_{\rm click}^{b_1,c_1d_1,b_2,\bar{c_2d_2}}=\sum_{n=0}^\infty \abs{p_n}^2\bigg\{\left[1-\frac{(1-t_{\rm B})\eta_{\rm det}}{2}\right]^n-2\left[1-\frac{\eta_{\rm det}}{2}\right]^n-\left[1-\frac{3(1-t_{\rm B})\eta_{\rm det}}{4}\right]^n+2\left[1-\frac{(3-t_{\rm B})\eta_{\rm det}}{4}\right]^n+\left[1-\frac{(1+t_{\rm B})\eta_{\rm det}}{2}\right]^n$\\
\quad\quad\quad\quad\quad\quad\ \ $-\left[1-\frac{(3+t_{\rm B})\eta_{\rm det}}{4}\right]^n\bigg\}$ \\
 \hline
$P_{\rm click}^{b_1,c_1d_1,{\bar b_2},c_2d_2}=\sum_{n=0}^\infty \abs{p_n}^2\bigg\{\left[1-\frac{t_{\rm B}\eta_{\rm det}}{2}\right]^n-\left[1-\frac{\eta_{\rm det}}{2}\right]^n-\left[1-\frac{(1+t_{\rm B})\eta_{\rm det}}{4}\right]^n+\left[1-\frac{(3-t_{\rm B})\eta_{\rm det}}{4}\right]^n-\left(1-t_{\rm B}\eta_{\rm det}\right)^n$\\
\quad\quad\quad\quad\quad\quad\ \ $+\left[1-\frac{(1+t_{\rm B})\eta_{\rm det}}{2}\right]^n +\left[1-\frac{(1+3t_{\rm B})\eta_{\rm det}}{4}\right]^n-\left[1-\frac{(3+t_{\rm B})\eta_{\rm det}}{4}\right]^n\bigg\}$ \\
  \hline
$P_{\rm click}^{{\bar b_1},{\bar{c_1d_1}},b_2,c_2d_2}=\sum_{n=0}^\infty \abs{p_n}^2\bigg\{\left[1-\frac{(1+t_{\rm B})\eta_{\rm det}}{4}\right]^n-\left[1-\frac{(3-t_{\rm B})\eta_{\rm det}}{4}\right]^n-\left[1-\frac{(1+3t_{\rm B})\eta_{\rm det}}{4}\right]^n+\left[1-\frac{(3+t_{\rm B})\eta_{\rm det}}{4}\right]^n\bigg\}$  \\
 \hline
$P_{\rm click}^{b_1,{\bar{c_1d_1}},b_2,c_2d_2}=\sum_{n=0}^\infty \abs{p_n}^2\bigg\{\left[1-\frac{(1-t_{\rm B})\eta_{\rm det}}{4}\right]^n-\left[1-\frac{3(1-t_{\rm B})\eta_{\rm det}}{4}\right]^n-2\left[1-\frac{(1+t_{\rm B})\eta_{\rm det}}{4}\right]^n+2\left[1-\frac{(3-t_{\rm B})\eta_{\rm det}}{4}\right]^n$\\
\quad\quad\quad\quad\quad\quad\ \ $+\left[1-\frac{(1+3t_{\rm B})\eta_{\rm det}}{4}\right]^n-\left[1-\frac{(3+t_{\rm B})\eta_{\rm det}}{4}\right]^n\bigg\}$  \\
\hline
$P_{\rm click}^{{\bar b_1},c_1d_1,b_2,c_2d_2}=\sum_{n=0}^\infty \abs{p_n}^2\bigg\{
\left[1-\frac{t_{\rm B}\eta_{\rm det}}{2}\right]^n-\left[1-\frac{\eta_{\rm det}}{2}\right]^n-\left[1-\frac{(1+t_{\rm B})\eta_{\rm det}}{4}\right]^n+\left[1-\frac{(3-t_{\rm B})\eta_{\rm det}}{4}\right]^n-\left(1-t_{\rm B}\eta_{\rm det}\right)^n$\\
\quad\quad\quad\quad\quad\quad\ \ $+\left[1-\frac{(1+t_{\rm B})\eta_{\rm det}}{2}\right]^n +\left[1-\frac{(1+3t_{\rm B})\eta_{\rm det}}{4}\right]^n-\left[1-\frac{(3+t_{\rm B})\eta_{\rm det}}{4}\right]^n\bigg\}$  \\
 \hline\hline
\end{tabular}
\caption{Values of the probabilities that appear in Eq.~(\ref{dieg}).
  }\label{table1nnnBB}
\end{table}
By inserting the values of these probabilities in Eq.~(\ref{dieg}), we find that $P_{\rm single-coin}$ can be expressed as
\begin{widetext}
\begin{eqnarray}\label{dieg2}
P_{\rm single-coin}&=&\sum_{n=0}^\infty \abs{p_n}^2\bigg\{-3\left[1-\frac{\eta_{\rm det}}{2}\right]^n+4\left[1-\frac{(3-t_{\rm B})\eta_{\rm det}}{4}\right]^n+2\left[1-\frac{(1+t_{\rm B})\eta_{\rm det}}{2}\right]^n-2\left[1-\frac{(3+t_{\rm B})\eta_{\rm det}}{4}\right]^n \nonumber \\
&-&3\left[1-\frac{(1+t_{\rm B})\eta_{\rm det}}{4}\right]^n+2\left[1-\frac{(1+3t_{\rm B})\eta_{\rm det}}{4}\right]^n+\left[1-\frac{(1-t_{\rm B})\eta_{\rm det}}{2}\right]^n-2\left[1-\frac{3(1-t_{\rm B})\eta_{\rm det}}{4}\right]^n \nonumber \\
&+&2\left[1-\frac{t_{\rm B}\eta_{\rm det}}{2}\right]^n-2\left(1-t_{\rm B}\eta_{\rm det}\right)^n+\left[1-\frac{(1-t_{\rm B})\eta_{\rm det}}{4}\right]^n\bigg\}.
\end{eqnarray}
\end{widetext}

On the other hand, we have that the parameter $P_{\rm double-coin}$ is given by the probability $P_{\rm click}^{b_1,c_1d_1,b_2,c_2d_2}$. This means that it can be written as
\begin{eqnarray}\label{Mon12}
P_{\rm double-coin}&=&{\rm Tr}\left({\tilde \rho}_{b_2b_1c_2c_1d_2d_1}M_{b_1,c_1d_1,b_2,c_2d_2}\right),\nonumber \\
\end{eqnarray}
 where the operator $M_{b_1,c_1d_1,b_2,c_2d_2}$ has the form
\begin{eqnarray}\label{Mon13}
M_{b_1,c_1d_1,b_2,c_2d_2}&=&\left(\openone_{b_2}-\ket{0}\bra{0}_{b_2}\right)\otimes\left(\openone_{b_1}-\ket{0}\bra{0}_{b_1}\right)\nonumber \\&\otimes&\left(\openone_{c_1d_1}-\ket{00}\bra{00}_{c1d_1}\right) \nonumber \\
&\otimes&\left(\openone_{c_2d_2}-\ket{00}\bra{00}_{c_2d_2}\right).
\end{eqnarray}
That is, this operator corresponds to having a ``click'' in the modes $b_2$, $c_2$ and/or $d_2$, $b_1$, and $c_1$ and/or $d_1$. By combining Eqs.~(\ref{Mon6})-(\ref{Mon12})-(\ref{Mon13}), we obtain
\begin{widetext}
\begin{eqnarray}\label{zam3}
P_{\rm double-coin}&=&1+\sum_{n=0}^\infty \abs{p_n}^2\bigg\{2\left[1-\frac{\eta_{\rm det}}{2}\right]^n-2\left[1-\frac{(3-t_{\rm B})\eta_{\rm det}}{4}\right]^n-\left[1-\frac{(1+t_{\rm B})\eta_{\rm det}}{2}\right]^n+\left[1-\frac{(3+t_{\rm B})\eta_{\rm det}}{4}\right]^n \nonumber \\
&+&2\left[1-\frac{(1+t_{\rm B})\eta_{\rm det}}{4}\right]^n-\left[1-\frac{(1+3t_{\rm B})\eta_{\rm det}}{4}\right]^n-\left[1-\frac{(1-t_{\rm B})\eta_{\rm det}}{2}\right]^n+\left[1-\frac{3(1-t_{\rm B})\eta_{\rm det}}{4}\right]^n \nonumber \\
&-&2\left[1-\frac{t_{\rm B}\eta_{\rm det}}{2}\right]^n+\left(1-t_{\rm B}\eta_{\rm det}\right)^n-\left[1-\frac{(1-t_{\rm B})\eta_{\rm det}}{4}\right]^n\bigg\}.
\end{eqnarray}
\end{widetext}

Finally, from Eqs.~(\ref{zam2})-(\ref{dieg2})-(\ref{zam3}), we obtain that $P_{\rm coin}^{\rm double}$ is given by Eq.~(\ref{resB}).

\section{Parameters $p_{\rm click}(k)$ and $p_{\rm coin}(k)$}\label{parque2}

In this Appendix, we provide the value for $p_{\tau}(k)$, with $\tau\in\{{\rm click}, {\rm coin}\}$, that we obtain after combining Eqs.~(\ref{resA})-(\ref{resB})-(\ref{as7}), 
\begin{widetext}
\begin{eqnarray}\label{as6}
p_{\rm click}(k)&=&(k-1)\Bigg\{1-\frac{1}{2}\sum_{{\it n}=0}^\infty \Bigg[\abs{{\it q}_{\it n}}^2\left(1-t_{\rm B}\eta_{\rm det}\right)^{\it n}+\abs{p_n}^2\left(1-\frac{t_{\rm B}\eta_{\rm det}}{2}\right)^n\Bigg]\Bigg\}, \nonumber \\
p_{\rm coin}(k)&=&(k-1)\Bigg\{1-\frac{1}{2}\sum_{{\it n}=0}^\infty \Bigg[\abs{q_n}^2\Big\{\left[1-(1-t_{\rm B})\frac{\eta_{\rm det}}{2}\right]^n+\left(1-\eta_{\rm det}t_{\rm B}\right)^n-\left[1-(1+t_{\rm B})\frac{\eta_{\rm det}}{2}\right]^n \Big\} \nonumber \\
&-&\frac{\abs{p_n}^2}{2}\bigg\{\left[1-\frac{\eta_{\rm det}}{2}\right]^n-2\left[1-\frac{t_{\rm B}\eta_{\rm det}}{2}\right]^n+\left[1-\frac{(1+t_{\rm B})\eta_{\rm det}}{4}\right]^n-\left[1-\frac{(1-t_{\rm B})\eta_{\rm det}}{2}\right]^n  \nonumber \\
&-&\left[1-\frac{(1-t_{\rm B})\eta_{\rm det}}{4}\right]^n\bigg\}\Bigg]\Bigg\},
\end{eqnarray}
\end{widetext}
for $k\geq{}2$.

\section{Expected Gain and coincidence detection rate}\label{exp}

In this Appendix, we calculate the expected gain and coincidence detection rate at Bob's side in the absence of Eve's attack. We denote these two expected values by $G$ and $G_{\rm coin}$, respectively. For this, we use a simple channel model.

Precisely, we consider a lossy quantum channel, and we disregard any misalignment effect both in the channel and in Alice's and Bob's apparatuses. Also, as already mentioned, we neglect the effect of the dark count probability of Bob's detectors. In particular, let $\eta_{\rm channel}=10^{-\alpha_{\rm channel} L/10}$ denote the transmittance of the quantum channel, where $\alpha_{\rm channel}$ ($L$) represents the loss coefficient (length) of the channel measured in dB/km (km). Also, we define $\eta_{\rm sys}= \eta_{\rm channel}\eta_{\rm det}$, where $\eta_{\rm det}$ is the detection efficiency of Bob's detectors, which we assume is equal for all of them. 

Let us first calculate the gain $G$. We need to consider three cases, depending on the actual signal sent by Alice. If Alice sends Bob the signal $\ket{\varphi_0}$, he receives in his data line the state $\ket{0}\ket{\sqrt{\eta_{\rm sys}t_{\rm B}}\alpha}$. This means that he can only observe a ``click'' in that line in the first time instant, which happens with probability
\begin{eqnarray}
P_{\rm click, \ket{\varphi_0}}=1-\left|\bra{0}\sqrt{\eta_{\rm sys}t_{\rm B}}\alpha\rangle\right|^2=1-e^{-\eta_{\rm sys}t_{\rm B}\mu},\ \ \ 
\end{eqnarray}
where $\mu=|\alpha|^2$. Likewise, it is straightforward to show that the probability that Bob observes a ``click'' in his data line when Alice sends him the signal $\ket{\varphi_1}$ satisfies $P_{\rm click, \ket{\varphi_1}}=P_{\rm click, \ket{\varphi_0}}$. On the other hand, if Alice sends Bob the signal $\ket{\varphi_2}$, we have that Bob receives in his data line the signal  $\ket{\sqrt{\eta_{\rm sys}t_{\rm B}}\alpha}\ket{\sqrt{\eta_{\rm sys}t_{\rm B}}\alpha}$ and, therefore, he observes a ``click'' with probability
\begin{eqnarray}\label{zxcv2}
P_{\rm click, \ket{\varphi_2}}&=&1-\left|\bra{0}\sqrt{\eta_{\rm sys}t_{\rm B}}\alpha\rangle\right|^4=1-e^{-2\eta_{\rm sys}t_{\rm B}\mu}.\ \ \ \ \ \
\end{eqnarray}
In Eq.~(\ref{zxcv2}) we have taken into account that whenever Bob observes in his data line two detection ``clicks'' within the same signal, he randomly assigns a single ``click'' event to it.

Putting all together, we have that $G$ can be written as
\begin{eqnarray}\label{f3}
G&=&\sum_{j=0}^2 p_{\ket{\varphi_j}}P_{\rm click, \ket{\varphi_j}}\nonumber \\
&=&1-(1-f)e^{-\eta_{\rm sys}t_{\rm B}\mu}-fe^{-2\eta_{\rm sys}t_{\rm B}\mu},
\end{eqnarray}
where the probabilities $p_{\ket{\varphi_j}}$, with $j=0,1,2$, are given by Eq.~(\ref{zxcv1}). We note that the parameter $L_{\rm zero}$ considered in the main text corresponds to the value of $L$ such that $G=G_{\rm zero}$.

To obtain $G_{\rm coin}$, we need to evaluate six cases, which depend on the actual signal state sent by Alice and on the previous optical {\it pulse} that she sent to Bob. The probability that the previous optical pulse sent by Alice is in a state $\ket{0}$ ($\ket{\alpha}$), which we shall denote by $p_{\ket{0}}$ ($p_{\ket{\alpha}}$), is equal to $p_{\ket{\varphi_0}}$ ($1-p_{\ket{\varphi_0}}$). According to Eq.~(\ref{zxcv1}), this means that 
\begin{equation}
p_{\ket{0}}=\frac{1-f}{2}, \quad{\rm and}\quad
p_{\ket{\alpha}}=\frac{1+f}{2}.
\end{equation}

\begin{figure}
\centering 
\centerline{\includegraphics*[scale=0.5]{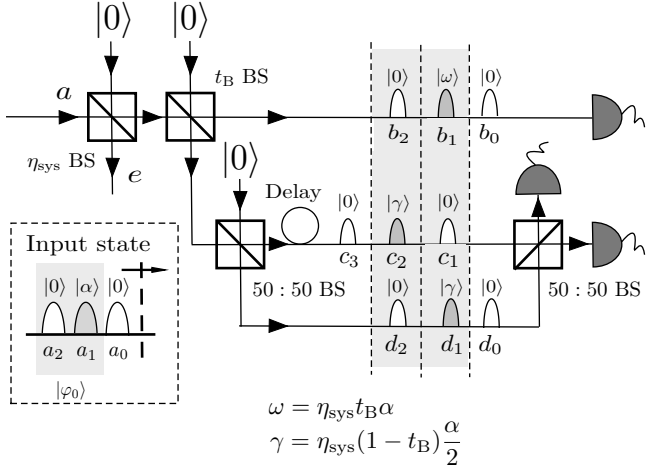}}
\caption{Schematic representation of the signals at Bob's receiver when Alice sends him the state $\ket{\varphi_0}=\ket{0}\ket{\alpha}$ preceded by the optical pulse $\ket{0}$. The  loss introduced by the quantum channel, together with the effect of the finite detection efficiency of Bob's detectors, is modelled with a beamsplitter (BS) of transmittance $\eta_{\rm sys}$ located at the input of Bob's receiver. We are interested in the probability that Bob observes a coincidence detection event in the time instants ``1'' and/or ``2''. 
}
\label{fig:t1}
\end{figure}

Let us consider first the scenario where Alice sends Bob the signal $\ket{\varphi_0}$ preceded by the optical pulse $\ket{0}$, which happens with probability $p_{\ket{\varphi_0}}p_{\ket{0}}=(1-f)^2/4$. This scenario is illustrated in Fig.~\ref{fig:t1}.
That is, here Alice sends him the following three optical pulses $\ket{0}_{a_2}\ket{\alpha}_{a_1}\ket{0}_{a_0}$, where we have introduced the subscripts $a_i$ to explicitly indicate the input mode $a$ and the time instants $i$. We shall denote by $P_{\rm coin, \ket{\varphi_0}\ket{0}}$ the average number of coincidente detection events observed by Bob (in the relevant instants ``1'' and ``2'') in this scenario. This parameter can be written as
\begin{eqnarray}\label{as1}
P_{\rm coin, \ket{\varphi_0}\ket{0}}&=&P_{\rm coin, \ket{\varphi_0}\ket{0},1}+P_{\rm coin, \ket{\varphi_0}\ket{0},2}\nonumber \\
&+&2P_{\rm coin, \ket{\varphi_0}\ket{0},{\rm double}},
\end{eqnarray}
where $P_{\rm coin, \ket{\varphi_0}\ket{0},j}$, with $j=1,2$, represents the probability that Bob observes a coincidence detection event only in the time instant ``$j$'' and no coincidence detection event in the time instant ``$1+[(j-1)\oplus 1]$'', and $P_{\rm coin, \ket{\varphi_0}\ket{0},{\rm double}}$ denotes the probability that Bob observes a coincidence detection event both in the time instants ``1'' and ``2''. 

As shown in Fig.~\ref{fig:t1}, there is vacuum in the data line in the time instant ``2'' ({\it i.e.}, in the mode $b_2$). This means that no coincidence detection event is possible in that time instant, {\it i.e.}, 
\begin{equation}\label{as2}
P_{\rm coin, \ket{\varphi_0}\ket{0}, 2}=P_{\rm coin, \ket{\varphi_0}\ket{0},{\rm double}}=0.
\end{equation}
On the other hand, in the time instant ``1'' there is a coherent state $\ket{\sqrt{\eta_{\rm sys}t_{\rm B}}\alpha}_{b_1}$ in the data line, and a vacuum state $\ket{0}_{c_1}$ and a coherent state $\ket{\sqrt{\eta_{\rm sys}(1-t_{\rm B})/2}\alpha}_{d_1}$ in the two arms of the monitoring line's interferometer. This means that $P_{\rm coin, \ket{\varphi_0}\ket{0},1}$ is given by
\begin{eqnarray}\label{as3}
P_{\rm coin, \ket{\varphi_0}\ket{0}, 1}&=&\left[1-\left|\bra{0}\sqrt{\eta_{\rm sys}t_{\rm B}}\alpha\rangle\right|^2\right]\nonumber \\
&\times&\left[1-\left|\bra{0}\sqrt{\frac{\eta_{\rm sys}(1-t_{\rm B})}{2}}\alpha\rangle\right|^2\right] \nonumber \\
&=&\left(1-e^{-\eta_{\rm sys}t_{\rm B}\mu}\right)\left(1-e^{-\eta_{\rm sys}(1-t_{\rm B})\mu/2}\right). \nonumber \\
\end{eqnarray}

By combining Eqs.~(\ref{as1})-(\ref{as2}) we obtain that $P_{\rm coin, \ket{\varphi_0}\ket{0}}$ is also given by Eq.~(\ref{as3}).

The other five cases can be analyzed similarly and we omit the details here for simplicity. We find that
\begin{eqnarray}
P_{\rm coin, \ket{\varphi_0}\ket{\alpha}}&=&\frac{1}{2}P_{\rm coin, \ket{\varphi_2}\ket{\alpha}}=\left(1-e^{-\eta_{\rm sys}t_{\rm B}\mu}\right)\nonumber \\
&\times&\left(1-e^{-\eta_{\rm sys}(1-t_{\rm B})\mu}\right), \nonumber \\
P_{\rm coin, \ket{\varphi_1}\ket{0}}&=&P_{\rm coin, \ket{\varphi_1}\ket{\alpha}}=\left(1-e^{-\eta_{\rm sys}t_{\rm B}\mu}\right)\nonumber \\
&\times&\left(1-e^{-\eta_{\rm sys}(1-t_{\rm B})\mu/2}\right), \nonumber \\
P_{\rm coin, \ket{\varphi_2}\ket{0}}&=&\left(1-e^{-\eta_{\rm sys}t_{\rm B}\mu}\right)\Big[\left(1-e^{-\eta_{\rm sys}(1-t_{\rm B})\mu/2}\right)\nonumber \\
&+&\left(1-e^{-\eta_{\rm sys}(1-t_{\rm B})\mu}\right)\Big]
\end{eqnarray}

Finally, if we include the a priori probabilities associated to the different six cases, we have that
\begin{eqnarray}\label{f4}
G_{\rm coin}&=&\sum_{j=0}^2\sum_{l\in\{0,\alpha\}} p_{\ket{\varphi_j}}p_{\ket{l}}P_{\rm coin, \ket{\varphi_j}\ket{l}}=\frac{1-e^{-\eta_{\rm sys}t_{\rm B}\mu}}{4} \nonumber \\
&\times&\Big\{(1-f)(3+f)\left[1-e^{-\eta_{\rm sys}(1-t_{\rm B})\mu/2}\right]\nonumber \\
&+&(1+6f+f^2)\left[1-e^{-\eta_{\rm sys}(1-t_{\rm B})\mu}\right]\Big\}.
\end{eqnarray}

\section{Monitoring detection rates: Eve's USD measurement}\label{meas_fri}

In this Appendix, we consider Eve's USD measurement when Alice and Bob monitor the detection rates of the signals. This measurement has four possible outcomes. Three of them identify each of the signals $\ket{\varphi_i}$, with $i=0,1,2$, while the fourth outcome denotes an inconclusive result. 

Precisely, we shall consider a USD measurement that maximizes the following quantity,
\begin{equation}\label{fri2}
(1-\zeta)q_{{\rm s},\zeta}+\zeta{}q_{{\rm s},\zeta}^{\rm d},
\end{equation}
for a certain $\zeta\in[f,1]$. Here, $q_{{\rm s},\zeta}$ ($q_{{\rm s},\zeta}^{\rm d}$) represents the conditional probability that the result is conclusive when the input signal to the measurement is a key generation (decoy) signal. In Eq.~(\ref{fri2}), we use the fact that, for the optimal USD measurement, $q_{{\rm s},\zeta}$ is equal for both key generation signals, as they have the same {\it a priori} probability. 

If $\zeta=f$, we recover the solution provided by the optimal USD measurement in~\cite{zero_cow}, {\it i.e.} the one that maximizes the probability to deliver a conclusive result. The case $\zeta=1$ corresponds to the USD measurement that provides the highest possible value for $q_{{\rm s},\zeta}^{\rm d}$, which we denote by $q_{\rm s, max}^{\rm d}$. When $f<\zeta<1$, we are in an intermediate scenario where $q_{\rm s}^{\rm d}\leq{}q_{{\rm s},\zeta}^{\rm d}\leq{}q_{\rm s, max}^{\rm d}$, with $q_{\rm s}^{\rm d}$ referring to the optimal solution in~\cite{zero_cow}. That is, with the parameter $\zeta$ we can tune the probability to correctly identify a decoy signal up to its maximum allowed value. The drawback of using $\zeta\neq{}f$ is that the overall probability to obtain a conclusive result (see Eq.~(\ref{fri1}) below) decreases when compared to its maximum possible value.  

From~\cite{zero_cow}, it follows directly that the optimal values for $q_{{\rm s},\zeta}$ and $q_{{\rm s},\zeta}^{\rm d}$ as a function of $\zeta$, $\mu$ and the parameter $\xi=\zeta/[2(1-\zeta)]$, are given by 
\begin{equation}
q^{\rm s}_{{\rm s},\zeta}=1-\exp(-\mu)\quad {\rm and}\quad q^{\rm d}_{{\rm s},\zeta}=0,
\end{equation}
 if $\sqrt{\xi}\leq\exp(-\mu/2)$. On the other hand, we have 
 \begin{eqnarray}
 q^{\rm s}_{{\rm s},\zeta}&=&1+\exp(-\mu)-\exp(-\mu/2)\sqrt{2\zeta/(1-\zeta)},\nonumber \\
 q^{\rm d}_{{\rm s},\zeta}&=&1-\sqrt{\xi^{-1}}\exp(-\mu/2),
 \end{eqnarray}
 if $\sqrt{\xi}> \exp(-\mu/2)$ and $\cosh\left(\mu/2\right)\geq \sqrt{\xi}$. Finally, if $\sqrt{\xi}> \exp(-\mu/2)$ and $\cosh\left(\mu/2\right)< \sqrt{\xi}$, we obtain 
 \begin{equation}
q^{\rm s}_{{\rm s},\zeta}=0\quad{\rm and}\quad q^{\rm d}_{{\rm s},\zeta}=\tanh\left(\mu/2\right).
\end{equation}

Given $q_{{\rm s},\zeta}$ and $q_{{\rm s},\zeta}^{\rm d}$, the probability $p_{{\rm c},\zeta}$ that Eve obtains a conclusive result with her USD measurement is then given by 
\begin{equation}\label{fri1}
p_{{\rm c},\zeta}=(1-f)q_{{\rm s},\zeta}+fq_{{\rm s},\zeta}^{\rm d}.
\end{equation}

\section{Expected Gain $G_{\rm decoy}$}\label{exp2}

Here, we calculate the expected gain of the decoy signals $\ket{\varphi_2}$ at Bob's data line in the absence of Eve's attack. For this, we use the simple channel model introduced in Appendix~\ref{exp}.

In particular, from Eqs.~(\ref{zxcv1})-(\ref{zxcv2})-(\ref{f3}) we have that
\begin{equation}
G_{\rm decoy}=p_{\ket{\varphi_2}}P_{\rm click, \ket{\varphi_2}}=f\left(1-e^{-2\eta_{\rm sys}t_{\rm B}\mu}\right).
\end{equation}

\section{Optimal USD measurement}\label{opt_usd}

In this Appendix, we calculate Eve's optimal USD measurement to discriminate the four signal states, $\ket{\varphi_i}$ with $i=0,\ldots, 3$, sent by Alice in the four-state COW-QKD protocol considered in Sec.~\ref{four}. 

This measurement is described by a positive-operator-valued measure (POVM) with five elements $E_i\geq0$, with $i=0,\ldots,4$, that satisfy $\sum_{i=0}^4 E_i=\openone$, with $\openone$ being the identity operator. $E_i$ identifies the signal $\ket{\varphi_i}$, with $i=0,\ldots, 3$, while $E_4$ corresponds to an inconclusive result. That is, we have that $p_{j|i}=\bra{\varphi_i}E_j\ket{\varphi_i}=0$  $\forall i\neq j$ with $i,j=0,\ldots,3$, with $p_{j|i}$ being the conditional probability of obtaining a result associated to the operator $E_j$ when measuring the state $\ket{\varphi_i}$. Moreover, since the signals $\ket{\varphi_0}$ and $\ket{\varphi_1}$ are sent with the same a priori probability, we impose $p_{0|0}=p_{1|1}$. This is illustrated in Table~\ref{table1n}.
\begin{table}[h!]
  \centering
  \begin{tabular}{lccccc}
    \quad & \multicolumn{4}{c}{Eve's POVM elements}\\
    \hline\hline
  Alice's signal & \quad $E_0$ \quad & \quad $E_1$ \quad & \quad $E_2$ \quad & \ \quad $E_3$  \quad & \ \quad $E_4$ \ \quad \\
  \hline
\quad\quad$\ket{\varphi_0}$   & \quad $q_{\rm s}$ \quad & \quad 0  \quad  & \quad 0  \quad  & 0 & $q_{\rm inc}$   \quad  \\
\quad\quad$\ket{\varphi_1}$   & \quad 0 \quad & \quad $q_{\rm s}$ \quad  & \quad 0 \quad  & 0 & $q_{\rm inc}$ \quad  \\
\quad\quad$\ket{\varphi_2}$   & \quad 0 \quad & \quad 0 \quad  & \quad $q_{\rm s}^{\rm d}$ \quad  & 0 & $q_{\rm inc}^{\rm d}$  \quad  \\ 
\quad\quad$\ket{\varphi_3}$   & \quad 0 \quad & \quad 0 \quad  & \quad 0 \quad  & $q_{\rm s}^{\rm v}$ & $q_{\rm inc}^{\rm v}$  \quad  \\ 
 \hline\hline
\end{tabular}
\caption{Conditional probabilities associated to Eve's optimal USD measurement. $E_4$ corresponds to an inconclusive result. Here, we use the following notation: $p_{0|0}=p_{1|1}=q_{\rm s}$, $p_{4|0}=p_{4|1}=q_{\rm inc}$, $p_{2|2}=q_{\rm s}^{\rm d}$, $p_{4|2}=q_{\rm inc}^{\rm d}$, $p_{3|3}=q_{\rm s}^{\rm v}$ and $p_{4|3}=q_{\rm inc}^{\rm v}$.
  }\label{table1n}
\end{table}

This means that the probability that Eve obtains a conclusive measurement result when she measures Alice's signals one by one with the USD measurement described above is given by
\begin{equation}\label{qs_qf_qinc}
p_{\rm c}=\sum_{i=0}^3 p_{\ket{\varphi_i}}p_{i|i}=(1-f_{\rm d}-f_{\rm v})q_{\rm s}+f_{\rm d}q^{\rm d}_{\rm s}+f_{\rm v}q^{\rm v}_{\rm s},
\end{equation}
while the probability to obtain an inconclusive result is $p_{\rm inc}=1-p_{\rm c}$. In Eq.~(\ref{qs_qf_qinc}) we use the notation introduced in Table~\ref{table1n} for the probabilities $p_{i|i}$. 

We find, therefore, that Eve's optimal USD measurement, {\it i.e.}, the one that maximizes $p_{\rm c}$, can be obtained by solving the following semidefinite program (SDP):
\begin{eqnarray}\label{sdp_eq}
&& {\rm maximize}\quad (1-f_{\rm d}-f_{\rm v})p_{0|0}+f_{\rm d}p_{2|2}+f_{\rm v}p_{3|3} \nonumber \\
&& {\rm s.\ t.}\quad \sum_{j=0}^4 E_j=\openone,\ E_j\geq0\ \ \forall j=0,\ldots,4, \nonumber \\
&& \quad\quad\,\,\,\,\, p_{j|i}=0\quad  \forall i,j=0,\ldots,3,\ {\rm with} \ i\neq j, \nonumber \\
&& \quad\quad\,\,\,\,\, p_{0|0}=p_{1|1}. 
\end{eqnarray}
Next, we explain how to solve the SDP given by Eq.~(\ref{sdp_eq}) numerically. 

\subsection{Solving the SDP numerically}\label{nume}

For this, we first rewrite Eq.~(\ref{sdp_eq}) in a more convenient way. Precisely, we express Alice's signals $\ket{\varphi_i}$ in some orthonormal basis $\{\ket{b_j}\}_{j=0,\ldots,3}$ of a four-dimensional Hilbert space $\mathcal{H}_4$,
\begin{eqnarray}\label{coef}
\ket{\varphi_0}&=&\ket{b_0},\nonumber \\
\ket{\varphi_1}&=&{e^{ - {{\left| \alpha  \right|}^2}}}\ket{b_0} + \sqrt {1 - {e^{ - 2{{\left| \alpha  \right|}^2}}}} \ket{b_1},\nonumber\\
\ket{\varphi_2}&=&{e^{ - \frac{{{{\left| \alpha  \right|}^2}}}{2}}}\ket{b_0} + {e^{ - \frac{{{{\left| \alpha  \right|}^2}}}{2}}}\frac{{1 - {e^{ - {{\left| \alpha  \right|}^2}}}}}{{\sqrt {1 - {e^{ - 2{{\left| \alpha  \right|}^2}}}} }}\ket{b_1}\nonumber \\
&+& \frac{{1 - {e^{ - {{\left| \alpha  \right|}^2}}}}}{{\sqrt {1 - {e^{ - 2{{\left| \alpha  \right|}^2}}}} }}\ket{b_2}, \nonumber\\
\ket{\varphi_3}&=&{e^{ - \frac{{{{\left| \alpha  \right|}^2}}}{2}}}\ket{b_0} + {e^{ - \frac{{{{\left| \alpha  \right|}^2}}}{2}}}\frac{{1 - {e^{ - {{\left| \alpha  \right|}^2}}}}}{{\sqrt {1 - {e^{ - 2{{\left| \alpha  \right|}^2}}}} }}\ket{b_1}\nonumber \\
&-&e^{-{\left| \alpha  \right|}^2} \frac{{1 - {e^{ - {{\left| \alpha  \right|}^2}}}}}{{\sqrt {1 - {e^{ - 2{{\left| \alpha  \right|}^2}}}} }}\ket{b_2}+\left(1-e^{-{\left| \alpha  \right|}^2}\right)\ket{b_3}.\nonumber \\
\end{eqnarray}
This implies that we can restrict ourselves to operators $E_j$ that act on $\mathcal{H}_4$.

Next, we rewrite the states $\rho_i=\ket{\varphi_i}\bra{\varphi_i}$ and the operators $E_j$ in terms of the generalized Gell-Mann matrices in $\mathcal{H}_4$. In general, say in $\mathcal{H}_n$, these matrices are $n\times n$ matrices $\sigma_k$, with $k=0\ldots, n^2-1$, that satisfy: $\sigma_0=\openone_n$ is the identity operator in $\mathcal{H}_n$, and the remaining $\sigma_k$ are traceless Hermitian matrices fulfilling 
\begin{eqnarray}\label{cond_m}
{\rm Tr}(\sigma_k)&=&n\delta_{k0}, \nonumber \\
{\rm Tr}(\sigma_k\sigma_l)&=&n\delta_{kl},
\end{eqnarray}
where $\delta_{kl}$ represents the Kronecker delta. 

The matrices $\sigma_k$ can be obtained by using three different types of matrices. First, we take $C_n^2=n!/[2!(n-2)!]$ symmetric matrices with all its elements equal to zero except the $p$th row $q$th column element and the $q$th row $p$th column element that are equal to $\sqrt{n/2}$, with $1\leq p< q\leq n$. Then, we take $C_n^2$ antisymmetric matrices with all its elements equal to zero except the $p$th row $q$th column element that is equal to $-i\sqrt{n/2}$ and the $q$th row $p$th column element that is equal to $i\sqrt{n/2}$, with $1\leq p< q\leq n$. Finally, we take $n-1$ diagonal matrices, which, for convenience, we denote by $\Delta^{(d)}$, with $1\leq d\leq n-1$, that satisfy
\begin{eqnarray}
\Delta^{(1)} &=& \sqrt{\frac{n}{2C_2^2}}
\begin{pmatrix}
1 & 0 & 0& \cdots & 0 \\
0 & -1 & 0& \cdots & 0 \\
0 & 0 & 0& \cdots & 0 \\
\vdots  & \vdots & \vdots  & \ddots & \vdots  \\
0 & \cdots & \cdots & \cdots & 0 
\end{pmatrix}, \nonumber \\
\quad \nonumber \\
\Delta^{(2)} &=& \sqrt{\frac{n}{2C_3^2}}
\begin{pmatrix}
1 & 0 & 0& \cdots & 0 \\
0 & 1 & 0& \cdots & 0 \\
0 & 0 & -2& \cdots & 0 \\
\vdots  & \vdots & \vdots  & \ddots & \vdots  \\
0 & \cdots & \cdots & \cdots & 0 
\end{pmatrix}, \quad  \nonumber \\
\quad \nonumber \\
\vdots & & \quad\quad\quad\quad\quad\quad\quad\quad\vdots
\quad \nonumber \\
\Delta^{(n-1)} &=& \sqrt{\frac{n}{2C_n^2}}
\begin{pmatrix}
1 & 0 & 0& \cdots & 0 \\
0 & 1 & 0& \cdots & 0 \\
0 & 0 & 1& \cdots & 0 \\
\vdots  & \vdots & \vdots  & \ddots & \vdots  \\
0 & \cdots & \cdots & \cdots & -(n-1)
\end{pmatrix}.\ \ \ \ \ 
\end{eqnarray}
In doing so, we indeed obtain $C_n^2+C_n^2+(n-1)=n^2-1$ matrices $\sigma_k$, with $k=1\ldots, n^2-1$, that satisfy Eq.~(\ref{cond_m}).

By using these matrices, we can rewrite $\rho_i$ and $E_j$ as follows
\begin{eqnarray}\label{sun98}
\rho_i&=&\sum_{k=0}^{n^2-1}\varphi_{ik}\sigma_k, \quad {\rm and}\quad 
E_j=\sum_{k=0}^{n^2-1}e_{jk}\sigma_k, \ \ \ \ \ 
\end{eqnarray}
for certain {\it known} real coefficients $\varphi_{ik}$, and for certain {\it unknown} real coefficients $e_{jk}$. 

For simplicity, we omit here the explicit values of the coefficients $\varphi_{ik}$. They can be obtained directly from~Eq.~(\ref{coef}) by using the fact that 
\begin{equation}
\varphi_{ik}=\frac{1}{n}{\rm Tr}(\rho_i\sigma_k). 
\end{equation}

This means, in particular, that the conditional probabilities $p_{j|i}$ can now be expressed as
\begin{equation}\label{cond_prob_sdp}
p_{j|i}={\rm Tr}(E_j\rho_i)=n\sum_{k=0}^{n^2-1}\varphi _{ik}e_{jk},
\end{equation}
where we have used Eqs.~(\ref{cond_m})-(\ref{sun98}).

Putting all together, and taking into account that $n=4$, we find that the SDP given by Eq.~(\ref{sdp_eq}) can be written as
\begin{eqnarray}\label{sdp_eq_v2}
&& {\rm maximize}\quad 4\sum_{k=0}^{15}\Big[(1-f_{\rm d}-f_{\rm vac})\varphi _{0k}e_{0k}+f_{\rm d}\varphi _{2k}e_{2k}\nonumber \\
&&\quad\quad\quad\quad\quad+f_{\rm vac}\varphi _{3k}e_{3k}\Big] \nonumber \\
&& {\rm s.\ t.}\quad \sum_{k=0}^{15}\sum_{j=0}^4e_{jk}\sigma_k-\sigma_0=0,\nonumber \\ 
&& \quad\quad\,\,\,\,\, \sum_{k=0}^{15}e_{jk}\sigma_k\geq0\ \ \forall j=0,\ldots,4, \nonumber \\
&& \quad\quad\,\,\,\,\, \sum_{k=0}^{15}\varphi _{ik}e_{jk}=0\quad  \forall i,j=0,\ldots,3,\ {\rm with} \ i\neq j, \nonumber \\
&& \quad\quad\,\,\,\,\, \sum_{k=0}^{15}
(\varphi _{0k}e_{0k}-\varphi _{1k}e_{1k})=0, 
\end{eqnarray}
where the unknown parameters are, as already mentioned, the coefficients $e_{jk}$. In this format, the SDP can be readily solved numerically to obtain the conditional probabilities $q_{\rm s}$, $q_{\rm inc}$, $q_{\rm s}^{\rm d}$, $q_{\rm inc}^{\rm d}$, $q_{\rm s}^{\rm v}$ and $q_{\rm inc}^{\rm v}$ given by Table~\ref{table1n} and, thus, also obtain the conclusive probability $p_{\rm c}$ given by Eq.~(\ref{qs_qf_qinc}). For this, we use the solver Mosek~\cite{mosek} and the input tool YALMIP~\cite{yalmip}.


\begin{thebibliography}{99}

\bibitem{qkd1} H.-K. Lo, M. Curty, and K. Tamaki, Secure quantum key distribution, Nat. Photonics {\bf 8}, 595 (2014).
\bibitem{qkd2} F. Xu, X. Ma, Q. Zhang, H.-K. Lo, and J.-W. Pan, Secure quantum key distribution with realistic devices, Rev. Mod. Phys. {\bf 92}, 025002 (2020).
\bibitem{qkd3} S. Pirandola {\it et al.}, Advances in Quantum Cryptography, Adv. Opt. Photon. {\bf 12}, 1012-1236 (2020).
\bibitem{pad} G. S. Vernam, Cipher printing telegraph systems for secret wire and radio telegraphic communications, J. Am. Inst. Electr. Eng. {\bf 45}, 109 (1926).
\bibitem{prop1} C. H. Bennett, and G. Brassard, in {\it Proceedings of the IEEE International Conference on Computers, Systems and Signal Processing}, (IEEE Press, Bangalore, India New York, 1984), p. 175.
\bibitem{prop2} A. K. Ekert, Quantum cryptography based on Bell's theorem, Phys. Rev. Lett. {\bf 67}, 661 (1991).
\bibitem{prot1} C. H. Bennett, Quantum cryptography using any two nonorthogonal states, Phys. Rev. Lett. {\bf 68}, 3121 (1992).
\bibitem{prot1b} K. Inoue, E. Waks, and Y. Yamamoto, Differential Phase Shift Quantum Key Distribution, Phys. Rev. Lett. {\bf 89}, 037902 (2002).
\bibitem{prot1c} T. Sasaki, Y. Yamamoto, and M. Koashi, Practical quantum key distribution protocol without monitoring signal disturbance, Nature {\bf 509}, 475 (2014).
\bibitem{cow1} N. Gisin, G. Ribordy, H. Zbinden, D. Stucki, N. Brunner, and V. Scarani, Towards practical and fast Quantum Cryptography, preprint arXiv:quant-ph/0411022 (2004). 
\bibitem{prot2} W.-Y. Hwang, Quantum Key Distribution with High Loss: Toward Global Secure Communication, Phys. Rev. Lett. {\bf 91}, 057901 (2003).
\bibitem{prot3} H.-K. Lo, X. Ma, and K. Chen, Decoy State Quantum Key Distribution, Phys. Rev. Lett. {\bf 94}, 230504 (2005).
\bibitem{prot4} X.-B. Wang, Beating the Photon-Number-Splitting Attack in Practical Quantum Cryptography, Phys. Rev. Lett. {\bf 94}, 230503 (2005).
\bibitem{prot5} M. Koashi, Unconditional Security of Coherent-State Quantum Key Distribution with a Strong Phase-Reference Pulse, Phys. Rev. Lett. {\bf 93}, 120501 (2004).
\bibitem{prot6} K. Tamaki, N. L\"utkenhaus, M. Koashi, and J. Batuwantudawe, Unconditional security of the Bennett 1992 quantum-key-distribution scheme with a strong reference pulse, Phys. Rev. A {\bf 80}, 032302 (2009).
\bibitem{prot7} X. Ma, C.-H. F. Fung, and H.-K.Lo, Quantum key distribution with entangled photon sources, Phys. Rev. A {\bf 76}, 012307 (2007).
\bibitem{prot8} D. Mayers, and A. Yao, in {\it Proceedings of the 39th Annual Symposium on Foundations of Computer Science}, (IEEE Computer Society, Los Alamitos, California, 1998), p. 503.
\bibitem{prot9} A. Ac\'in, N. Brunner, N. Gisin, S. Massar, S. Pironio, and V. Scarani, Device-Independent Security of Quantum Cryptography against Collective Attacks, Phys. Rev. Lett. {\bf 98}, 230501 (2007).
\bibitem{prot10} H.-K. Lo, M. Curty, and B. Qi, Measurement-Device-Independent Quantum Key Distribution, Phys. Rev. Lett. {\bf 108}, 130503 (2012).
\bibitem{prot11} M. Lucamarini, Z. Yuan, J. Dynes, and A. Shields, Overcoming the rate-distance limit of quantum key distribution without quantum repeaters, Nature {\bf 557}, 400 (2018). 
\bibitem{prot12} X.-B. Wang, Z.-W. Yu, and X.-L. Hu, Twin-field quantum key distribution with large misalignment error, Phys. Rev. A {\bf 98}, 062323 (2018).
\bibitem{prot13} M. Curty, K. Azuma, and H.-K. Lo, Simple security proof of twin-field type quantum key distribution protocol, npj Quantum Inf. {\bf 5}, 64 (2019).
\bibitem{impl1} Y. Zhao, B. Qi, X. Ma, H.-K. Lo, and L. Qian, Experimental Quantum Key Distribution with Decoy States, Phys. Rev. Lett. {\bf 96}, 070502 (2006).
\bibitem{impl2} C.-Z. Peng, J. Zhang, D. Yang, W.-B. Gao, H.-X. Ma, H. Yin, H.-P. Zeng, T. Yang, X.-B. Wang, and J.-W. Pan, Experimental Long-Distance Decoy-State Quantum Key Distribution Based on Polarization Encoding, Phys. Rev. Lett. {\bf 98}, 010505 (2007).
\bibitem{impl3} D. Rosenberg, J. W. Harrington, P. R. Rice, P. A. Hiskett, C. G. Peterson, R. J. Hughes, A. E. Lita, S. W. Nam, and J. E. Nordholt, Long-Distance Decoy-State Quantum Key Distribution in Optical Fiber, Phys. Rev. Lett. {\bf 98}, 010503 (2007).
\bibitem{impl4} A. R. Dixon, Z. L. Yuan, J. F. Dynes, A. W. Sharpe, and A. J. Shields, Gigahertz decoy quantum key distribution with 1 Mbit/s secure key rate, Opt. Express {\bf 16}, 18790 (2008).
\bibitem{impl5} A. Poppe {\it et al.}, Practical quantum key distribution with polarization entangled photons, Opt. Express {\bf 12}, 3865-3871 (2004).
\bibitem{impl6} A. Treiber, A. Poppe, M. Hentschel, D. Ferrini, T. Lor\"unser, E. Querasser, T. Matyus, H. H\"ubel, and A. Zeilinger, A fully automated entanglement-based quantum cryptography system for telecom fiber networks, New J. Phys. {\bf 11}, 045013 (2009).
\bibitem{impl7} H. Takesue, S. W. Nam, Q. Zhang, R. H. Hadfield, T. Honjo, K. Tamaki, and Y. Yamamoto, Quantum key distribution over a 40-dB channel loss using superconducting single-photon detectors, Nat. Photonics {\bf 1}, 343 (2007).
\bibitem{cow2} D. Stucki, N. Brunner, N. Gisin, V. Scarani, and H. Zbinden, Fast and simple one-way quantum key distribution, Appl. Phys. Lett. {\bf 87}, 194108 (2005).
\bibitem{impl9} D. Stucki, N. Walenta, F. Vannel, R. T. Thew, N. Gisin, H. Zbinden, S. Gray, C. R. Towery, and S. Ten, High rate, long-distance quantum key distribution over 250 km of ultra low loss fibres, New J. Phys. {\bf 11}, 075003 (2009).
\bibitem{cow4} B. Korzh, C. C. W. Lim, R. Houlmann, N. Gisin, M. J. Li, D. Nolan, B. Sanguinetti, R. Thew, and H. Zbinden, Provably secure and practical quantum key distribution over 307 km of optical fibre, Nat. Photonics {\bf 9}, 163 (2015).
\bibitem{impl10} A. Rubenok, J. A. Slater, P. Chan, I. Lucio-Martinez, and W. Tittel, Real-World Two-Photon Interference and Proof-of-Principle Quantum Key Distribution Immune to Detector Attacks, Phys. Rev. Lett. {\bf 111}, 130501 (2013).
\bibitem{impl11} T. Ferreira da Silva, D. Vitoreti, G. B. Xavier, G. C. do Amaral, G. P. Tempor\~ao, and J. P. von der Weid, Proof-of-principle demonstration of measurement-device-independent quantum key distribution using polarization qubits, Phys. Rev. A {\bf 88}, 052303 (2013).
\bibitem{impl12} Y. Liu {\it et al.}, Experimental Measurement-Device-Independent Quantum Key Distribution, Phys. Rev. Lett. {\bf 111}, 130502 (2013).
\bibitem{impl13} A. Boaron {\it et al.}, Secure Quantum Key Distribution over 421 km of Optical Fiber, Phys. Rev. Lett. {\bf 121}, 190502 (2018).
\bibitem{net1} D. Stucki, M. Legr\'e, F. Buntschu, B. Clausen, N. Felber, N. Gisin, L. Henzen, P. Junod, G. Litzistorf, and P. Monbaron, Long-term performance of the SwissQuantum quantum key distribution network in a field environment, New J. Phys. {\bf 13}, 123001 (2011).
\bibitem{net2} M. Sasaki {\it et al.}, Field test of quantum key distribution in the Tokyo QKD Network, Opt. Express {\bf 19}, 10387 (2011).
\bibitem{net3} J. Qiu, Quantum communications leap out of the lab, Nature {\bf 508}, 441 (2014).
\bibitem{net4} J. Dynes {\it et al.}, Cambridge quantum network, npj Quantum Information {\bf 5}, 1 (2019).
\bibitem{impl14} M. Pittaluga {\it et al.}, 600 km repeater-like quantum communications with dual-band stabilisation, preprint arXiv:2012.15099 (2020).
\bibitem{sat1} S.-K. Liao {\it et al.}, Satellite-to-ground quantum key distribution, Nature {\bf 549}, 43 (2017).
\bibitem{sat2} H. Takenaka, A. Carrasco-Casado, M. Fujiwara, M. Kitamura, M. Sasaki, and M. Toyoshima, Satellite-to-ground quantum-limited communication using a 50-kg-class microsatellite, Nat. Photon. {\bf 11}, 502 (2017). 
\bibitem{scale1} M. Takeoka, S. Guha, and M. M. Wilde, Fundamental rate-loss tradeoff for optical quantum key distribution, Nat. Commun. {\bf 5}, 5235 (2014).
\bibitem{scale2} S. Pirandola, R. Laurenza, C. Ottaviani, and L. Banchi, Fundamental limits of repeaterless quantum communications, Nat. Commun. {\bf 8}, 15043 (2017).
\bibitem{company} ID Quantique, Geneva, Switzerland, http://www.idquantique.com.
\bibitem{upper} J. Gonz\'alez-Payo, R. Tr\'enyi, W. Wang, and M. Curty, Upper Security Bounds for Coherent-One-Way Quantum Key Distribution, Phys. Rev. Lett. {\bf 125}, 260510 (2020).
\bibitem{zero_cow} R. Tr\'enyi, and M. Curty, Zero-error attack against coherent-one-way quantum key distribution, preprint arXiv:2101.07192 (2021).
\bibitem{low_cow} T. Moroder, M. Curty, C. C. W. Lim, L. P. Thinh, H. Zbinden, and N. Gisin, Security of Distributed-Phase-Reference Quantum Key Distribution, Phys. Rev. Lett. {\bf 109}, 260501 (2012).
\bibitem{cow_zer} C. Branciard, N. Gisin, N. L\"utkenhaus, and V. Scarani, Zero-Error Attacks and Detection Statistics in the Coherent One-Way Protocol for Quantum Cryptography, Quant. Inf. Comput. {\bf 7}, 639 (2007).
\bibitem{cow_zer1} D. A. Kronberg, A. S. Nikolaeva, Y. V. Kurochkin, and A. K. Fedorov, Quantum soft filtering for the improved security analysis of the coherent one-way quantum-key-distribution protocol, Phys. Rev. A {\bf 101}, 032334 (2020).
\bibitem{chefles_usd1} A. Chefles, Unambiguous discrimination between linearly independent quantum states, Phys. Lett. A {\bf 239}, 339 (1998).
\bibitem{chefles_usd2} A. Chefles, and S. M. Barnett, Optimum unambiguous discrimination between linearly independent symmetric states, Phys. Lett. A {\bf 250}, 223 (1998).
\bibitem{eldar1} Y. C. Eldar, A semidefinite programming approach to optimal unambiguous discrimination of quantum states, IEEE Trans. Inform. Theory {\bf 49}, 446 (2003).
\bibitem{condition} M. Curty, M. Lewenstein, and Norbert L\"utkenhaus, Entanglement as a Precondition for Secure Quantum Key Distribution, Phys. Rev. Lett. {\bf 92}, 217903 (2004).
\bibitem{foot2} This corresponds to the case where Eq.~(\ref{sun3}) holds, which is the parameter regime we are interested in, as this condition is satisfied by the experimental implementations of COW-QKD. 
\bibitem{mosek} MOSEK ApS, Copenhagen, Denmark, https://www.mosek.com.
\bibitem{yalmip} J. L\"ofberg, in {\it Proceedings of the CACSD Conference} (Taipei, Taiwan, 2004), pp. 284-289, available from https://yalmip.github.io.
\bibitem{foot} Obviously, $\ket{\varphi_3}=\ket{\varphi_{\rm vac}}$. However, we prefer to use two different notations for the same signal because it allows us to distinguish the cases where Alice sent the state $\ket{\varphi_3}$ and Eve correctly identified this state (in which case we use $\ket{\varphi_3}$), from those instances where Eve obtained an inconclusive result with her measurement and she replaced the original signal sent by Alice with two vacuum optical pulses (in which case we use $\ket{\varphi_{\rm vac}}$).

\end{thebibliography}
\end{document}